\documentclass[onecolumn,a4paper,times,11pt]{aastex631}

\usepackage[varg]{txfonts}%
\usepackage{graphicx, natbib}%
\usepackage{enumerate}%
\usepackage{etoolbox}%
\bibpunct{(}{)}{,}{a}{}{,}%
\usepackage{textcomp}
\usepackage{gensymb}
\usepackage{enumitem}
\usepackage{hyperref}
\setlist{nosep} 

\newcommand{\kms}{km~s$^{-1}$}

\newcommand{\myvec}[1]{\ensuremath{\mathbf{#1}}}
\apptocmd{\lim}{\limits}{}{}

\usepackage{hyperref}
\usepackage{xcolor, soul}
\sethlcolor{yellow}

\begin{document}
\title{On general correlation between 3D solar wind speed and density model and solar proxies}
\author[0000-0001-8252-4104]{C. Porowski}
\affil{Space Research Centre PAS (CBK PAN),\\
Bartycka 18A, 00-716 Warsaw, Poland}
\author[0000-0003-3957-2359]{M. Bzowski}
\affil{Space Research Centre PAS (CBK PAN),\\
Bartycka 18A, 00-716 Warsaw, Poland}
\author[0000-0002-2982-1887]{M. Tokumaru}
\affil{Institute for Space-Earth Environmental research, Nagoya University, Nagoya, Japan}
\date{\today}
\keywords{
Fast solar wind (1872), Slow solar wind (1873), Solar wind (1534), Solar activity (1475), Stellar activity (1580), Heliosphere (711), Astrosphere interstellar medium interactions (106)
}

\begin{abstract}
The solar wind (SW) is a supersonic outflow of plasma from the solar corona, with the latitudinal speed and density profiles varying with the solar activity. The SW protons charge exchange with the inflowing interstellar neutral atoms and create energetic neutral atoms (ENAs), which bring information on the physical state of the plasma within the boundary region of the heliosphere. The speed of the ENAs depends on their energies, and consequently observations at different energies provide information on different epochs backwards in time. Therefore, understanding the history of the evolution of the SW is important to understand this information. In this paper, we extend the work by \citet{porowski_etal:22a}, who provided the WawHelioIon 3DSW model of the time evolution of latitudinal profiles of the SW speed and density based on results of analysis of interplanetary scintillations (IPS). Based on results of Principal Component Analysis, we seek for correlation between selected solar proxies and the structure of the SW obtained from IPS and show that it is possible to reproduce the evolution of the SW structure during the past three solar cycles based on the proxies. With this, we extend the history of the evolution of the SW structure back to 1976, i.e., to the epoch when observations of the key proxies -- the inclination of the SW current sheet and the solar polar magnetic fields -- became available. We point out the potential of the use of the proxies for forecasting the structure of the SW into the future.
\end{abstract}


\section{Introduction}
\label{sec:intro}
\noindent
The heliosphere is a bubble in the interstellar medium blown by the solar wind (SW). Information on processes operating in the boundary region between the SW and the interstellar gas can be obtained at 1 au from observations of energetic neutral atoms (ENAs) of hydrogen that are created by charge exchange between interstellar neutral atoms and PUI protons carried by the SW \citep{gruntman:97,gruntman_etal:01a}. The latitudinal structure of the SW varies during the cycle of solar activity, with the velocity and density profiles being relatively flat during the activity maxima and highly structured during the minima \citep{bzowski_etal:03a, lallement_etal:10b, mccomas_etal:08a, tokumaru_etal:10a}. The intensity of the production of hydrogen ENAs depends on the flux of interstellar PUIs, which depends on the flux and speed of the SW, how PUIs
are accelerated at the heliospheric termination shock and in the inner
heliosheath \citep[e.g.,][]{zank_etal:96b, chalov_etal:03a, yang_etal:15a, fahr_etal:16a, kumar_etal:18a, mostafavi_etal:19a, giacalone_etal:21a, zirnstein_etal:21c}, and
the physical processes governing the origin of the IBEX ribbon \citep[e.g.,][]{mccomas_etal:09c,heerikhuisen_etal:10a,schwadron_mccomas:10a,zirnstein_etal:15a, florinski_etal:16a,dayeh_etal:19a, mousavi_etal:22a}. The fluxes of ENAs observed at 1 au are additionally modulated by re-ionization losses due to charge-exchange with SW protons and photoionization inside the termination shock of the SW \citep{bzowski:08a}. 

The charge-exchange interaction of interstellar neutral hydrogen with the structured SW inside the heliosphere results in variations of the density and flux of interstellar hydrogen within a few au from the Sun \citep{bzowski_etal:02,sokol_etal:19b}, where most of the heliospheric backscatter glow observed at 1 au is formed \citep{rucinski_bzowski:95b}. Especially in the downwind hemisphere, including the region near the ecliptic plane, the latitudinal variation of the SW structure affects the density of interstellar neutral H and the helioglow intensity \citep{bzowski:03}. These variations are also visible in direct-sampling observations of interstellar neutral H atoms performed at 1 au by IBEX-Lo experiment onboard the Interstellar Boundary Explorer (IBEX) mission \citep{galli_etal:19a, rahmanifard_etal:19a}. The variations in the density of interstellar neutral H affect the production rate of the PUIs.

ENAs observed at 1 au at different energies bring information on the state of the plasma in the heliosheath at different epochs \citep{zirnstein_etal:15c, reisenfeld_etal:19a} because of the energy-dependent travel times of ENAs from their creation sites to the detector. Also the primary and secondary populations of interstellar neutral H feature long travel times, equal to several solar cycle periods \citep{bzowski_kubiak:20a}. With the advent of fully time-dependent models of the heliosphere, able to use models of time evolution of the solar factors \citep[e.g.,][]{izmodenov_alexashov:15a}, modeling time variations of the ENA productions becomes feasible.

Therefore, the evolution of the SW structure is important for interpretation of observations of heliospheric ENAs, interstellar neutral atoms, and the helioglow, in addition to the interest in the SW itself. Pioneering studies of the latitudinal structure of the SW were performed {\it in situ} by Ulysses during its three orbits in a solar polar plane. The SW speed and density profiles are available for one solar maximum and two solar minima \citep{wenzel_etal:89a, bame_etal:92a, gloeckler_etal:92, mccomas_etal:08a}. Especially valuable are those performed at the perihelion portions of the orbit, the so-called Fast Latitude Scans, where the full range of heliolatitudes was scanned within only several months. The measurements performed during the remaining portions of the Ulysses orbits are more challenging to interpret because the duration of the heliolatitude scan was comparable to the time scale of changes in the solar activity during the solar activity cycle. 

In practice, systematic monitoring of the SW structure is only possible using remote-sensing methods based on observations performed from the ecliptic plane. One of these is the interplanetary scintillation (IPS) method. It relies on multiple-station observations of scintillation of radio waves from compact radio sources, caused by density fluctuations of SW electrons \citep{hewish_etal:64a, manoharan:93b, kojima_kakinuma:90a}. The computer-aided tomography of IPS observations enable reliable determination of the solar wind structure inside 1 au \citep{jackson_etal:97a, jackson_etal:98a, kojima_etal:98a, tokumaru_etal:10a, tokumaru_etal:12b, tokumaru_etal:21a}. The SW speed is determined by observing IPS for a radio source using geographically-separated antennas. Based on observations of multiple sources performed during the period of solar rotation, a synoptic map of the solar wind speed as a function of heliolatitude and heliolongitude is produced. The SW speeds determined using the IPS method involve the effect of weighted integration along the line of sight. The tomographic analysis, in which correlation between the SW speed and the strength of density fluctuations is assumed, enables correcting for this effect. These monthly maps, released once per year, are publicly available from the Institute for Space-Earth Environmental Research in Nagoya, Japan \citep{tokumaru_etal:21a}. 

Inevitably, because of the geometry of the observations, the speeds obtained for the highest latitudes are the most uncertain \citep[see, e.g., Figure 7 in ][]{sokol_etal:13a}. Also, the coverage of the sky is very non-uniform because of the non-uniform distribution of suitable radio sources and the location of the antennas in the northern hemisphere of the Earth. Additionally, there are seasonal gaps in the sky coverage because of antenna maintenance during winter months. Individual data points in the synoptic maps feature a large scatter and some of them are outliers. However, heliospheric studies need a continuous and regular coverage, free from outlying elements. 

To address the postulate of a continuous and regular coverage, \citet{sokol_etal:13a} and \citet{bzowski_etal:13a} decided to sacrifice the longitudinal and a portion of the temporal resolution of the IPS SW speeds for the sake of regularity. They averaged the Carrington synoptic speed maps over heliolongitudes during individual calendar years, obtaining yearly heliolatitudinal speed profiles. Subsequently, they fitted parameters of an analytic model defined as first- and second-order smoothly-connected splines and obtained an analytic model of the yearly speed profiles, which can be linearly interpolated between individual years in time. This approach was subsequently continued by \citet{sokol_etal:19a} and \citet{sokol_etal:20a}. 

During the processing of the Carrington maps, \citet{sokol_etal:13a} observed that the magnitude of the speed was correlated with the number of line of sight observations used to compile the map. Consequently, they decided to discard the monthly maps developed based on the number of individual line of sight observations below a certain threshold. 

The presence of potential outliers in the Carrington maps of the SW speed and limitations of the analytic model used by \citet{sokol_etal:13a} and \citet{sokol_etal:20a} prompted \citet{porowski_etal:22a} to revisit the approach to the approximation of the SW structure. These authors developed a novel statistical method of identifying and rejecting outlying data in the Carrington maps. They cleaned the yearly latitudinal profiles of the SW speed and approximated them with a sum of the Legendre polynomials up to an order varying between the years, with an additional constraint of disappearing of the derivatives of the profiles over heliolatitude at the solar poles. 

Results of application of this scheme to reprocessed SW tomography data for the entire observation period 1987---2020, presented by \citet{tokumaru_etal:21a}, who modified the relation between the scintillation intensity and the SW speed, were collected in the WawHelioIon 3DSW model of the SW speed, almost free from outliers. The mathematical form of the yearly solar wind latitudinal speed profiles in the WawHelioIon 3DSW model is defined in Equations 1--2 in \citet{porowski_etal:22a}. A yearly profile is represented by a sum of the Legendre polynomials up to a certain order $n$, with the coefficients numbered from 0 to $n$. The model features an additional constraint of null derivative over heliolatitude at the poles, which reduces the number of independent parameters of the model to $n-1$. The parameters of the model make a time series covering the time interval 1985---2020. 

In the present paper, we continue the development of the WawHelioIon 3DSW model. We extend the measurement data base with observations from 2021. Because the time series created from the SW model coefficients covers almost four solar cycles, we expect that it is long enough to comprehend information on the temporal evolution of the SW on time scales longer than one solar activity cycle. We look for correlations between the shapes of the solar wind speed profiles, and demonstrate that a carefully selected set of proxies is able to reproduce the observed profiles of the solar wind speed. 

The objective of the paper is to find a general correlation between solar proxies and the SW structure that will provide a capability to calculate the mean yearly SW speed profile based on a selection of proxies. We examine the properties of the proxy time series to assess its applicability in expanding the SW evolution model in time (backward and forward).

The analysis starts with a mathematical formulation of the generalized SW model in Section \ref{sec:formula}. As a preparation for implementation of this model, 
in Section \ref{sec:swModelRearrang} we adapt the SW model developed by \citet{porowski_etal:22a} used as the input SW model, and in Section \ref{sec:dimensionReduction} we reduce its dimensionality using the Principal Components Analysis (PCA) method. Subsequently, in Section \ref{sec:proxySelection}, we take proxies that potentially can be used for reproduction of the SW speed profiles and, applying again the PCA analysis, we select a subset of proxies that is actually used to construct the generalized SW model. Next, in Section \ref{sec:extendedModel}, we present the extended SW model,
which we refer as the Generalized SW model based on selected proxies, and show its example application.
In the Discussion (Section \ref{sec:appliDiscu}), we show the performance of the proxy filtering and provide some comments on the model properties when applied to individual solar cycles.
We complete this section with a comparison of the proxy-based model with in situ SW speed observations from Ulysses (Section \ref{sec:ulyssess}). 
The results of the model development are presented in Section \ref{sec:extendedModel}, where all details needed to implement the model are given.
The paper is completed with a summary and conclusions in Section \ref{sec:Summary}. 

\section{Mathematical framework}
\label{sec:formula}
\noindent
The SW model recently published by \citet{porowski_etal:22a} provides a functional description of 35 yearly mean latitudinal profiles of the SW speed for the years 1985---2020. The profiles are obtained as a linear combinations of bases built from Legendre functions (the model base) and one of $m = 35$ individual vectors of parameters (the parameter vectors). Each of the parameter vectors is individually assigned to one of the 35 years in the range 1985---2020, without year 2010. The parameter vectors are obtained from separate fits of the base functions to IPS data. The parameter vectors, together with the base functions, provide a precise estimation of a yearly-averaged latitudinal SW speed profile for the middle of a given year.


In this work, we adopt a collection of solar proxies to see whether it is possible to express the SW model parameters through the proxies. Both the collection of solar proxies and the SW model parameters are treated as two potentially correlated multidimensional time series, in which we eliminate temporal dependence between them using a multilinear approach to express one by another. The SW model parameters used in this work are obtained from fits of the Legendre-base SW model to the IPS data that have been previously processed using the methodology developed by \citet{porowski_etal:22a}, with an extension to the data covering year 2021, which became available after completion of the analysis performed by these authors. The choice of proxies for the final model is based on an analysis that we discuss later in Sec. \ref{sec:proxySelection}.  

In the multilinear approach, it is desirable to provide uncorrelated inputs. Because both the commonly known proxies and the SW model parameters are usually highly correlated, we decompose them into uncorrelated components using PCA. Mathematical details of PCA are available in many textbooks, including, e.g., \citet{jolliffe:04a} and \citet{shlens:14a}. As a result of the application of PCA, the model parameters are not only decomposed into uncorrelated components, but also potential multilinearities and correlations between the model parameters are removed. Additionally, the PCA has also a feature of model dimensionality reduction, which is also used in our analysis.

The multilinear formulation we used may be expressed as follows. Denoting the $i$-th Principal Component (PC) of the input SW model parameter as $PC_i^{\text{SW model}}$ and the vector of proxy PC as $\myvec{PC}^{\text{proxy}}$, we assume a general relation between them in the form: 
\begin{equation}
    PC_i^{\text{SW model}}=c_i+\sum_{j=1}^n M_{ij}\,PC_j^{\text{proxy}},
\label{eq:YvsX}
\end{equation}
where $c_i$ is the $i$-th constant parameter, $M_{ij}$ is an element of an $m \times n$ matrix of coefficients \myvec{M}, which is describing a linear combination of the SW model parameters and the proxies, $n$ is the number of proxies, $PC_j^{\text{proxy}}$ is the $j$-th element of the proxy PC vector. In the application of this formula to the PCs of the parameters of the reduced model, the index $i$ ranges from 0 to $m$. Our objective is to determine the matrix \myvec{M} by means of parameter fitting for all years for the optimum values of $n$ and $m$. The values of $n$ and $m$ are found empirically.

In the formulation shown above, the matrix $\myvec{M}$ and the vector $\myvec{c}$ may be regarded as a generalization of the SW speed model, since the information included in matrix $\myvec{M}$ represents a generalized response of the model to a given proxy PC vector. 
Even though $\myvec{M}$ is retrieved using a proxy time series for a limited number of years, the time is not explicit in this formulation. 
So, under assumptions that (1) the general dependence between the proxies and SW profiles deduced from the fitting is a representative description of the dependence between the proxies and SW speed profile in general, i.e., for all times, and (2) that the transformation matrices used to calculate the PCs of the proxies and the PCs of the parameters also are representative, the matrix $\myvec{M}$ may be regarded as a generalized estimator of the SW profiles outside the time range covered by the available time series of the available IPS   observations, which returns SW speed profiles when a proxy vector is multiplied by this matrix.

The matrix $\myvec{M}$ and the vector $\myvec{c}$ are obtained by fitting to the data for the years 1985---2021, after necessary rearrangements. Since the components of $PC^{\text{SW model}}$ and $PC^{\text{proxy}}$ are uncorrelated, the fitting procedure that minimizes the expression in the time domain is performed separately for each component of $PC^{\text{SW model}}$. It means that each of the $m$ components, ($PC^{\text{SW model}}_i$) is fitted individually and expressed as a linear combination of the components of $PC^{\text{proxy}}$. We used a Wolfram Mathematica function NonlinearModelFit with automatic selection of minimization method. 

Similarly as in the case of the transformation matrix from the Legendre base to that obtained from the PCA analysis for the SW, the robustness of $\myvec{M}$ will be verified during the coming years, when more SW IPS data will be available. In our opinion, it is desirable to use the data from a time interval covering full solar cycles. At present, we have covered three full cycles and a small portion of the fourth one. We decided to not exclude the most recent years from the analysis to acknowledge for the presence of long-term variations in the solar wind, with a time scale much longer than that of the 22-year Hale cycle.
    
Using the above formulation, we performed an empirical study of the generalized SW model, which allowed us to determine the optimum model setup, which provides the best SW model estimations. The optimum setup that we developed brings stable and unique solutions of the SW profile estimations with a reasonable accuracy and satisfactory agreement with the Ulysses data. The setup determines the magnitudes of $m$ and $n$, i.e., the numbers of $m_{\text{model}}$ of the components of $PC^{\text{SW model}}$ and the $n_{\text{proxy}}$ components of $PC^{\text{proxy}}$, and the composition of the proxy vector. The quality of the solutions provided by the optimum model setup was checked by comparing the SW profile estimations provided by the generalized model for a wide variety of possible $m_{\text{model}}$ and $n_{\text{proxy}}$ values, as well as for different proxy vector compositions. More details will be presented below.

\section{Input SW model: its temporal properties and modifications}
\label{sec:swModelRearrang}
\subsection{Refitting of the SW Legendre model}
\noindent
As the input model we use the model based on Legendre functions from \citet{porowski_etal:22a}. In order to apply the formula from Equation \ref{eq:YvsX}, the input model is refitted to obtain a common order for all the years. 
The order of the input SW model varies from one year to another, resulting in a variable number of model parameters (i.e., the parameter vectors lengths) hanging from one year to another within the range between 9 and 17.
The variable order of the parameters was previously used to obtain a distribution of the residuals as close to the Gaussian as possible. It was motivated by the goal that the model is supposed to provide an optimal and precise SW speed profile, treating each year individually.
However, a constant dimension of the input model is required in the formula in Equation \ref{eq:YvsX}.
Therefore, the analysis starts with adoption of a common order of the SW input model, which requires refitting of the model. The use of a common order of the input SW model also allows us to apply the PCA in the analysis. The refitting includes also an extension of the IPS data by another year (2021), for which the IPS data are now available. 
This increases the number of years with data available to 36. 
The preprocessing of the IPS data for this additional year was done identically as for all the other years \citep{porowski_etal:22a}.

The first issue to address was selection of the common initial model order $m_{init}$. We found that the generalized SW model provides almost identical estimations of the SW profiles for a wide range of the $m_{init}$ values 8---22 when a filtration level at 1\% in the PC domain is adopted. We selected $m_{init}=20$,
which was motivated by the available length of the proxy vector limited by data and filtering. From this, we obtained 19 independent SW model parameters which are fitted to each of the 36 years individually. Each parameter corresponds to a given model order (and so to a given element of the Legendre base), so the fitting provides 19 vectors with 36 vlues of model parameters, which can be regarded as 19 sets of the time series.

\begin{figure}[!ht]
\center{
       \includegraphics[width=.95\columnwidth]{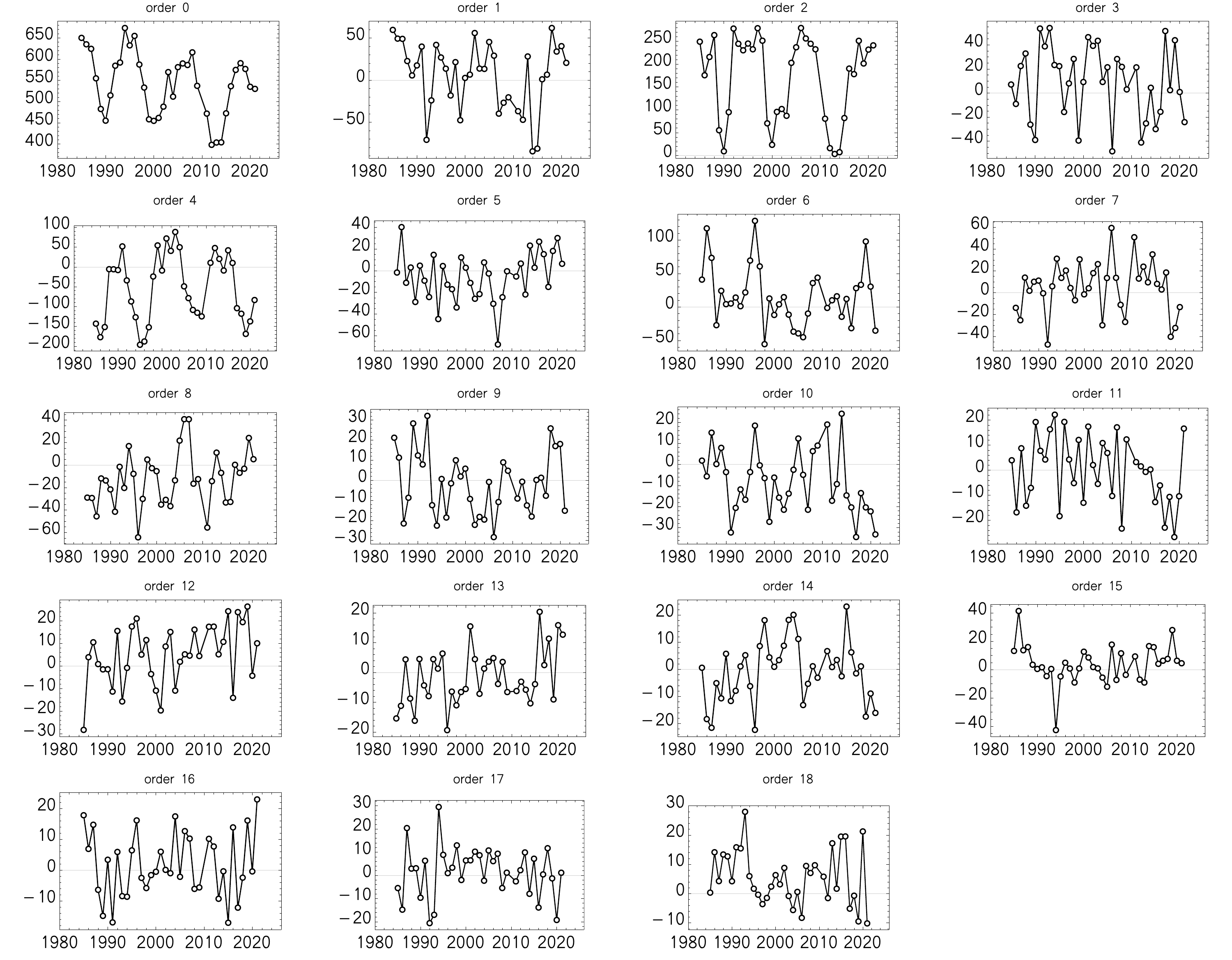}
	\caption{
		Time series of the parameters of the initial SW model for the years 1985---2021 obtained from fitting. 
				}
\label{fig:coeffs}
}
\end{figure}

Next, we analyze the time series composed of the refitted coefficients. The time series of the fit parameters of the SW model are shown in Figure \ref{fig:coeffs}.

\begin{figure}[!ht]
\center{
       \includegraphics[width=.95\columnwidth]{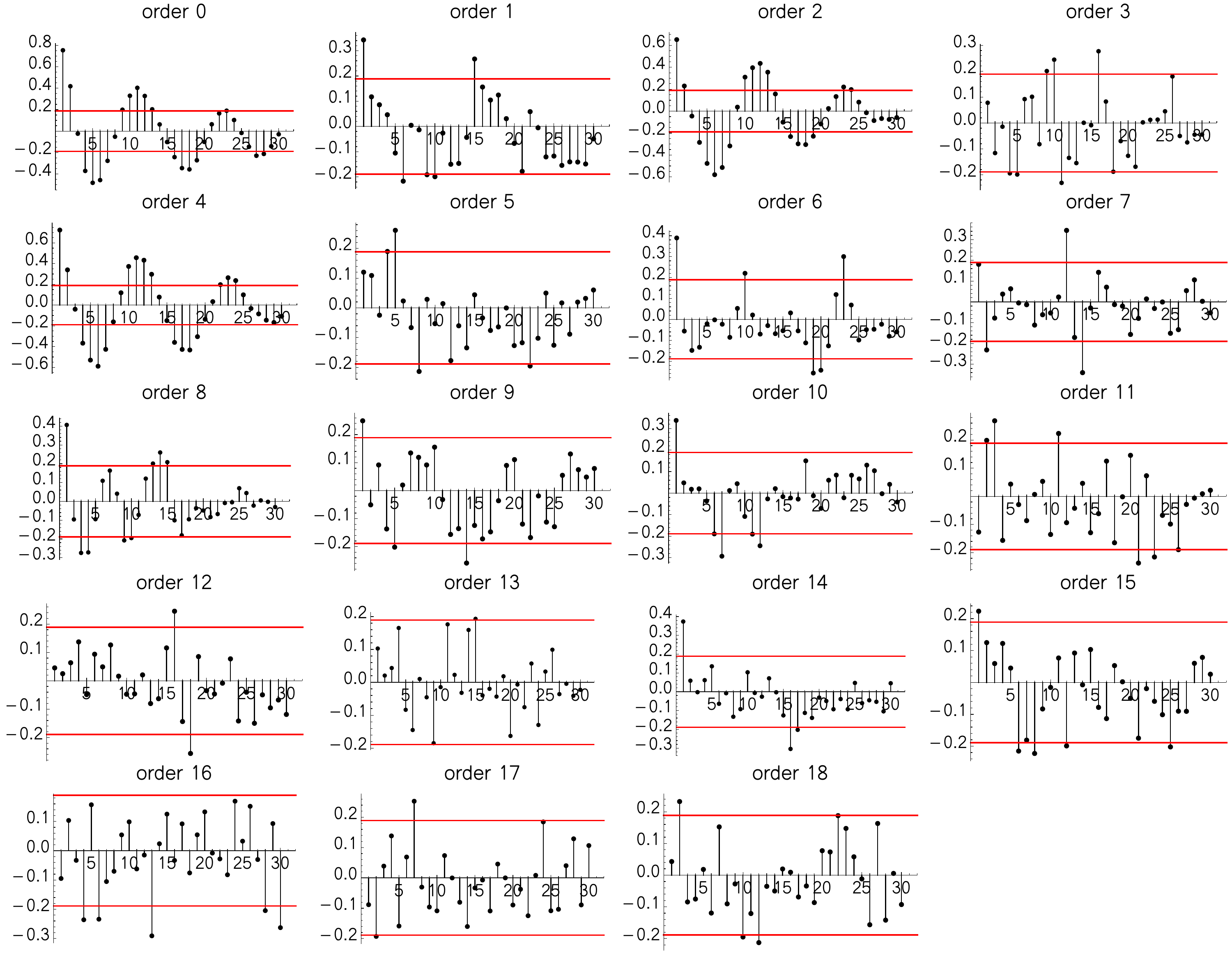}
    \caption{The magnitudes of the autocorrelation function for the 19 fit parameters of the refitted SW model, plotted as a function of the time lag expressed in years. The numbers above the panels indicate the orders corresponding to the Legendre base in the SW model. A 75\% confidence level bands are plotted with the red lines.
    }
\label{fig:ccsSymm}
}
\end{figure}

Visual inspection of the time series of the model parameters reveals a quasi-periodic behavior in some of them. In particular, the zeroth, second, and fourth parameters clearly show a periodicity corresponding to the solar cycle variations. While the zeroth parameter (i.e., a constant value) corresponds to the latitude-averaged SW speed, the physical and temporal interpretation of the other parameters is more complex. Therefore, to investigate their properties, we analyze the temporal variations of the parameters using the autocorrelation function (ACF), calculated for time lags ranging from 1 to 30. The results are plotted in Figure \ref{fig:ccsSymm} with the confidence bands for the confidence level $cl=75$\%.

Analysis of the ACF of the SW parameters confirms the existence of seasonality in the first three even-numbered model parameters, i.e., for \#0, \#2, and \#4, which describe the latitudinally-symmetric variation of the speed profiles and closely follow the 11 years solar cycle period. The remaining parameters feature periodicities close to 11 years (like \#6) or different than 11 years, being in general agreement with the SW periods reported by, e.g., \citet{prabhakaran:06a} and
\citet{katsavarias_etal:12a}. While the parameters \#1 and \#3 show the strongest periodicity for $\sim16$ years lag, a $\sim5.5$ years periodicity seems to be featured in the ACF of parameter \#5. The coefficient \#7 features a strong periodicity of $\sim 12.5$ years. In general, the odd-numbered parameters, which describe the asymmetric part of the model, show different properties than the even-numbered ones. No straightforward solar cycle-related seasonality is visible, but since their ACFs
show that the statistically significant lags are present and often close to the periods existing in the power spectrum of the SW speed, it is apparent that traces of a complex periodic behavior of the SW in the odd-numbered parameters exist, and that the SW model parameters are related to the physical parameters of the Sun. However, in this work we will not try to convey how the parameters correspond to the physical properties of the Sun, leaving it to be investigated in the future. It is also not excluded that at the moment the character of the temporal structures for the odd-numbered parameters may be not fully descriptive because of the limited statistics and because during the investigated time interval the SW featured secular (or very long-period) variations, including a drop in the mean flux \citep[see, e.g.,][]{mccomas_etal:13b, sokol_etal:21a}.

\subsection{Reduction of the dimensionality of the input SW model}
\label{sec:dimensionReduction}
\begin{figure}[!ht]
\center{
       \includegraphics[width=.75\columnwidth]{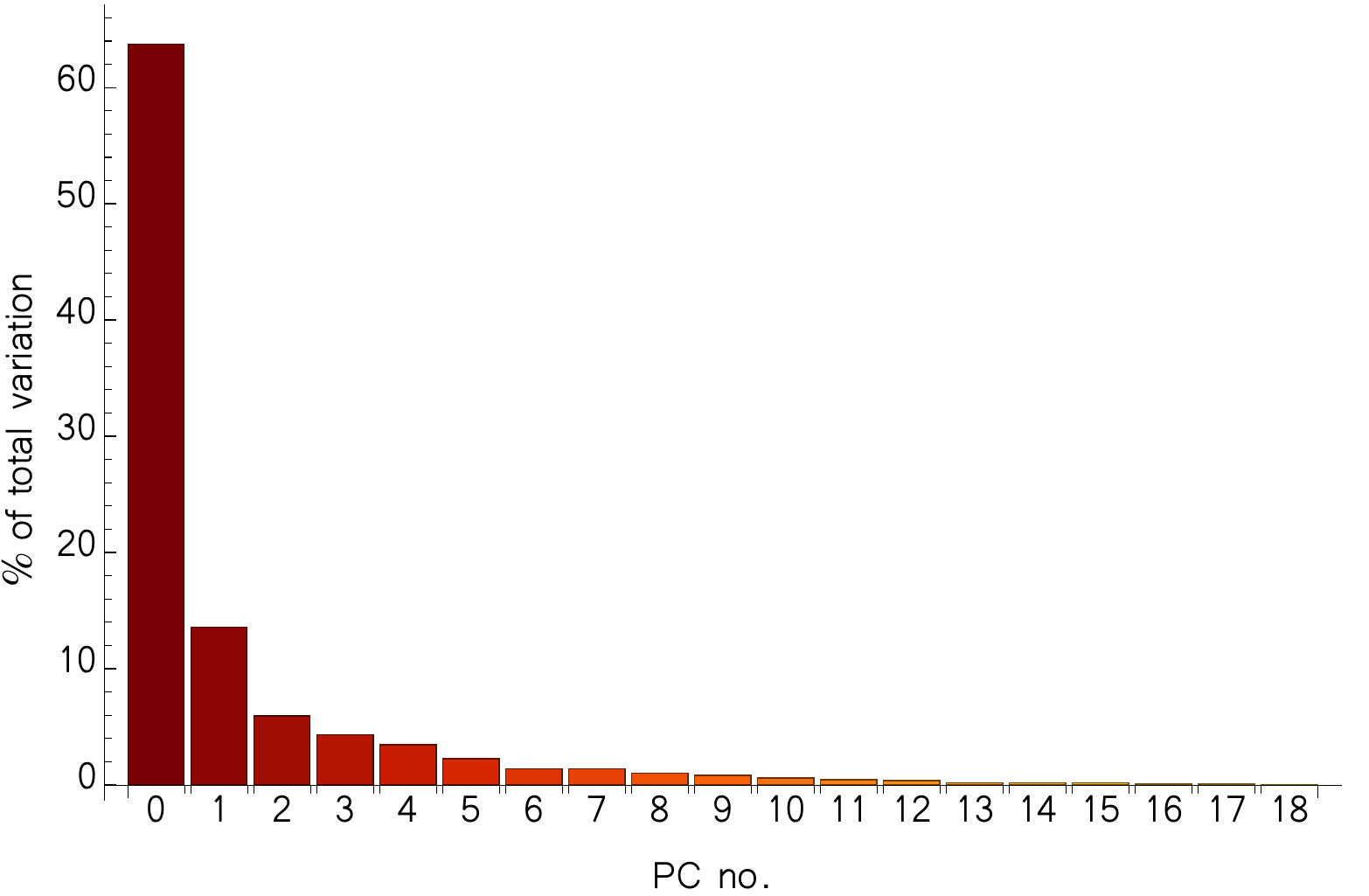}
	\caption{Contributions of individual PCs of the initial SW model to the total variation, expressed in percent.}
\label{fig:PCAbars}
}
\end{figure}

\begin{figure}[!ht]
\center{
        \includegraphics[width=.99\columnwidth]{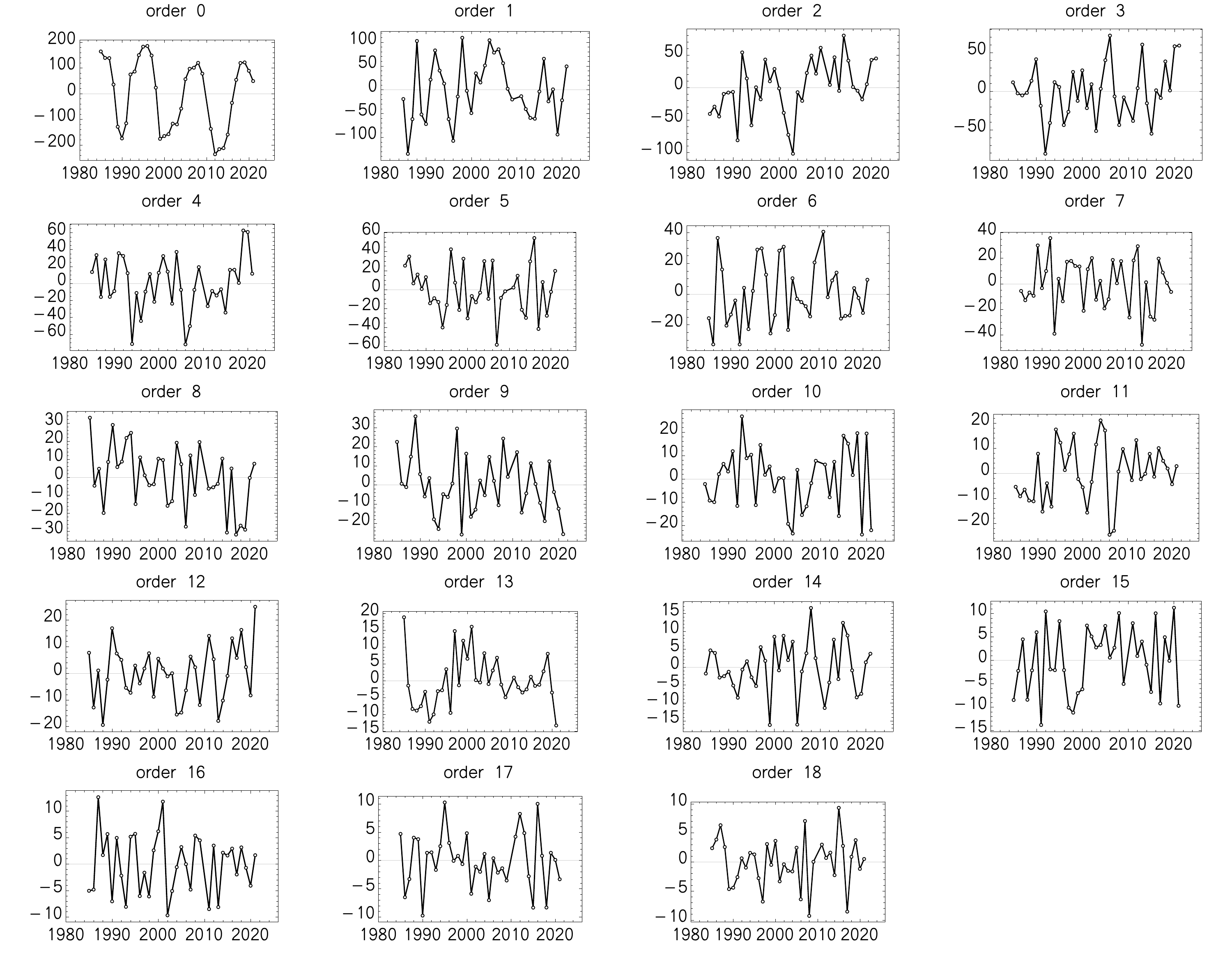}
    \caption{Time series of the 19 parameters of the SW initial model transformed into the PCA space.
    }
\label{fig:19PCA}
}
\end{figure}

\begin{figure}[ht]
\center{
       \includegraphics[width=.95\columnwidth]{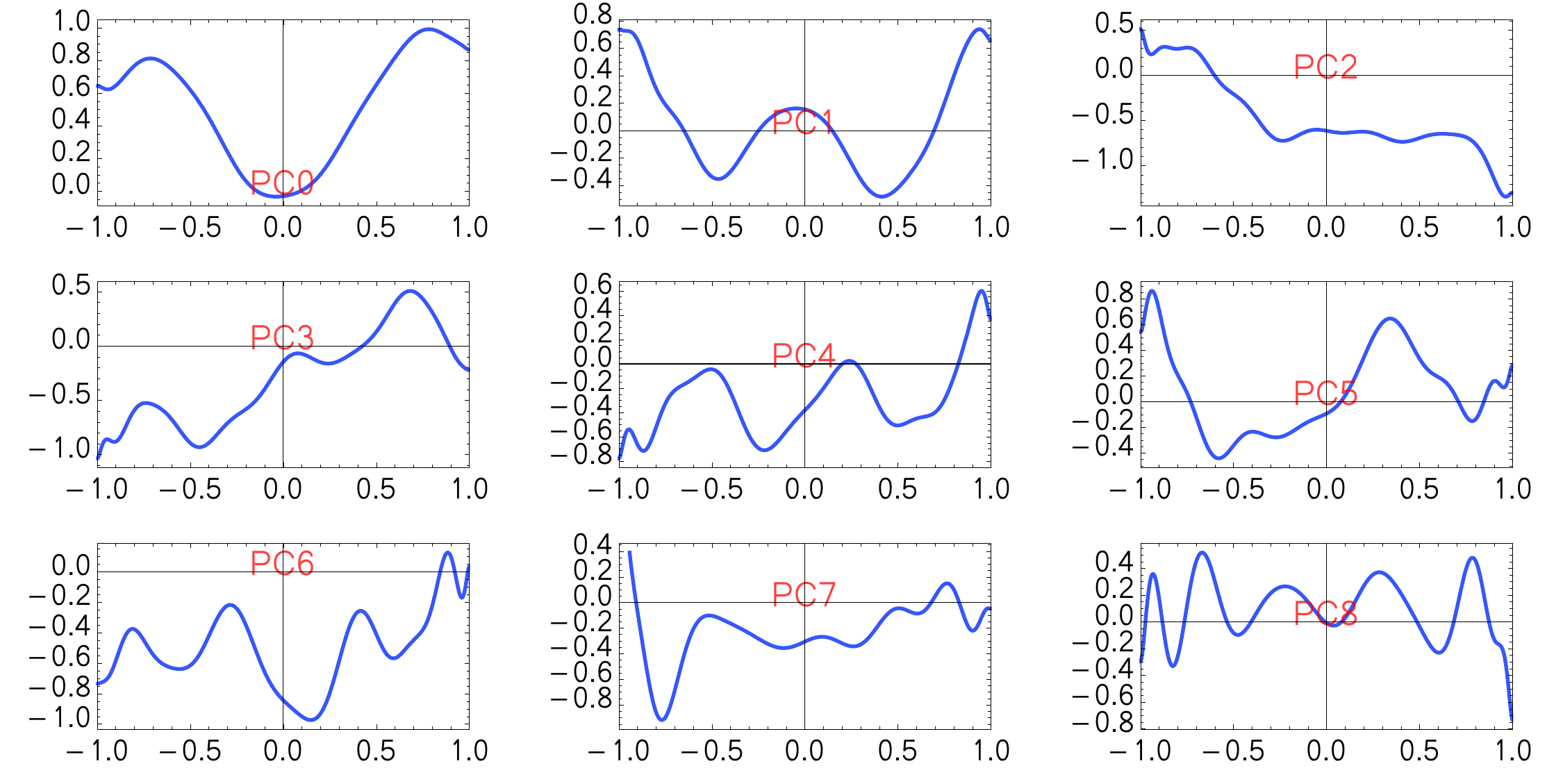}
       \caption{The base functions for the reduced SW model in the PC coordinate system.
       The base functions in this space are linear combinations of the original Legendre base functions of the SW model, obtained from the transformation matrix to the PCs coordinate system.
       Each of the base functions corresponds to an individual PC, as labeled in the plots.}
\label{fig:bases}
}
\end{figure}

\noindent
The common seasonality, seen in the ACF for some of the SW model parameters in Figure \ref{fig:ccsSymm}, strongly suggests that some of these parameters may be redundant. Therefore, as the next step of the analysis, we decompose the parameters into uncorrelated components using
PCA and filter out the least significant of the resulting components from further steps, i.e., we perform filtering in the PC space. A bar plot showing the contributions of the PCs of the SW model parameters to the total variation, ordered by their significance, is shown in Figure \ref{fig:PCAbars}. The time series of the SW model parameter PCs are shown in Figure \ref{fig:19PCA},
and a new base of the model, being a transformed Legendre bases into the PCA space, is shown in Fig. \ref{fig:bases}.

At this stage of research we feel unable to conclude if the transformation from the Legendre base to the PCA-optimized base that we have just derived has an universal character, or if acquiring more IPS observations, expected during the forthcoming yeas, will significantly modify the PCA-derived basis. We anticipate that at least a few first base functions will not change significantly, because we verified that for different model dimensionalities $n_{\text{init}}$ from the range 8---22 the first few bases are very similar. The parameters of the SW model in the PCA base are denoted $PC^{\text{SW model}}$. They may be regarded as the coefficients of the corresponding elements of the new base. 

The elements of the new base are weighted by the corresponding PCs. For example, it is seen that the first PC (PC0), which contributes to the total variation by more than 60\%, shows a seasonality that follows the heliolatitude-averaged SW speed, which modulates the corresponding V-like shaped element of the base (see the upper-left panel in Figure \ref{fig:bases}). Consequently, PC0 modulates the SW profile into the characteristic V-shape during the minima of the solar activity. The other components have a lower weight and contribute to details of the shapes of the yearly profiles. While PC1 modulates mostly the polar regions, PC2 and PC3 describe the north/south (N/S) asymmetry. 

\begin{figure}[!ht]
\center{
       \includegraphics[width=1.\columnwidth]{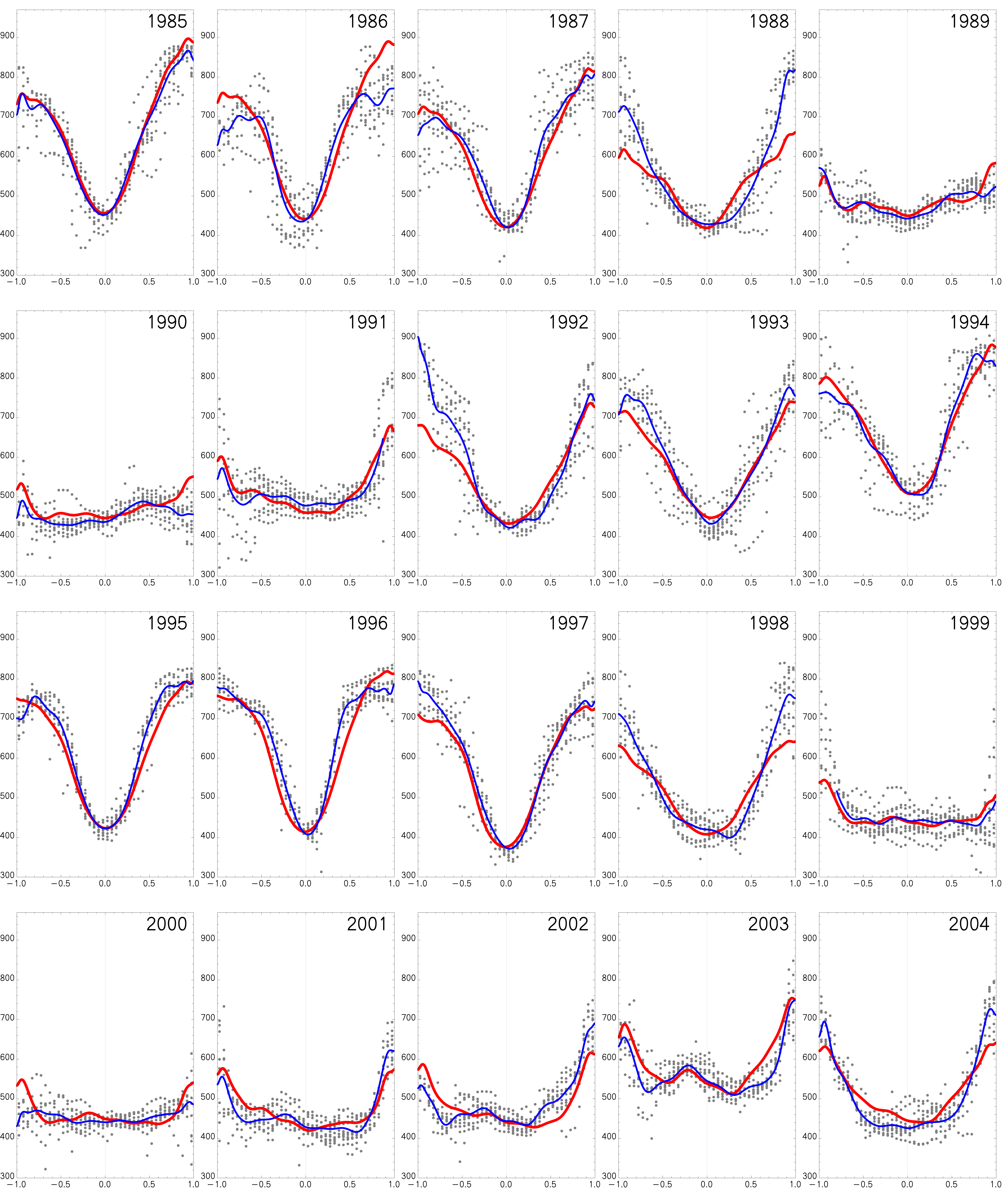}
        \caption{
		Solar wind speed data in \kms{} obtained from IPS analysis, filtered against outliers \citep[dots; ][]{porowski_etal:22a}, compared with the reduced SW models (blue lines) and with the generalized SW model (red lines). It is seen that despite the dimensionality of the generalized model is reduced, the SW profiles are almost identical to the initial model in most of the cases. The plots are in the $\cos(\phi)$ space.
	}
\label{fig:SWComp}
}
\end{figure}

\begin{figure}[!ht]
\center{
       \includegraphics[width=1.\columnwidth]{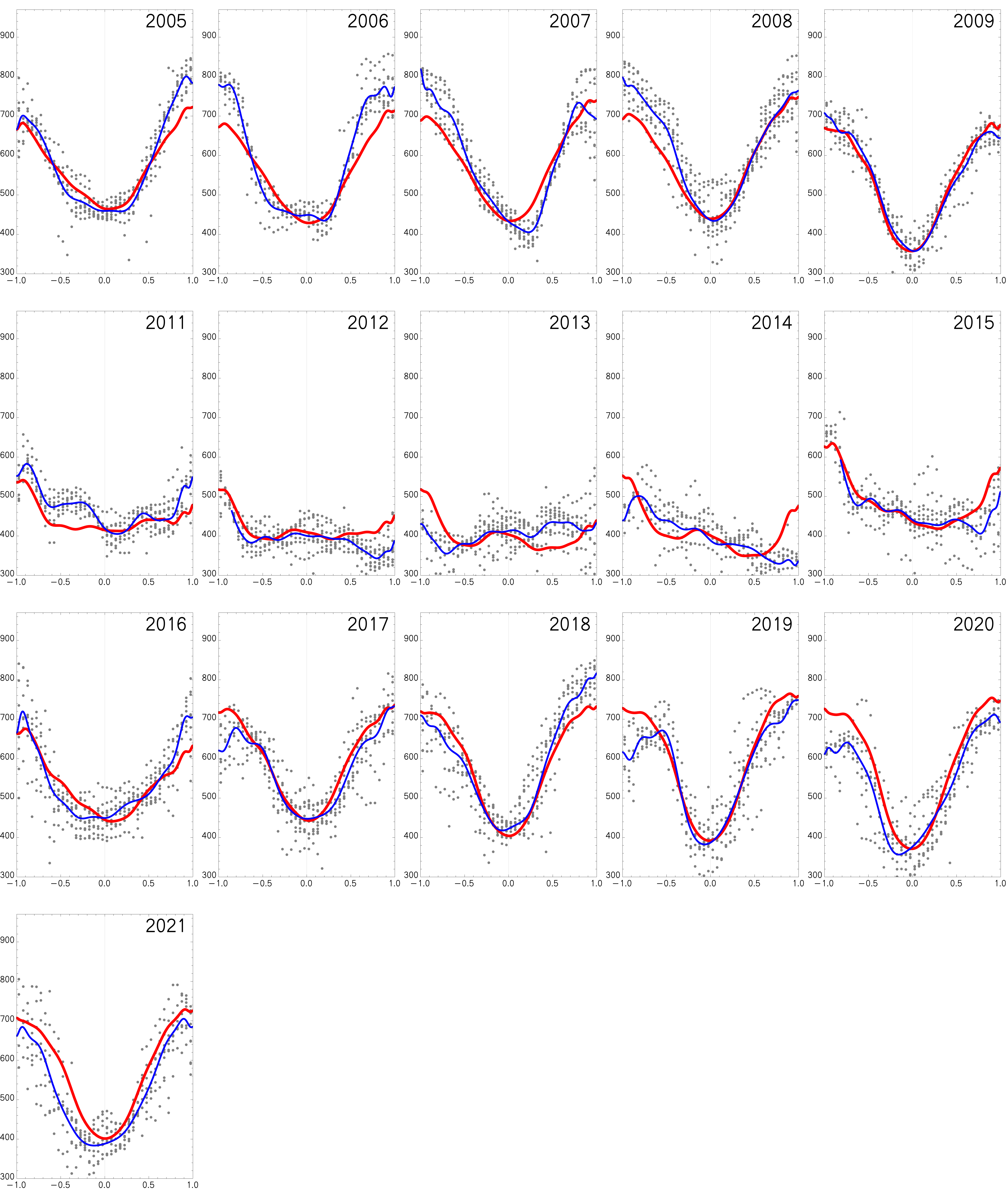}
	\caption{Figure \ref{fig:SWComp} continued.
	}
\label{fig:SWCompB}
}
\end{figure}

We determine the optimal order of the SW model assuming a certain cutoff level of the PC contribution to the total variance, below which the PCs are considered as insignificant and rejected. Based on empirical study we found that the optimum cutoff value for the PCs of the coefficients is 1\% of the contribution to the total variance. This reduces the number of the SW model parameters from 19 to 9. The model with the reduced common dimensionality will be referred to as the reduced SW model. The reduced SW model is used in fitting the proxy-based model. A comparison of the solar wind data, the original model from \citet{porowski_etal:22a} (extended to 2021), and the reduced model are presented in Figures \ref{fig:SWComp} and \ref{fig:SWCompB}.

\begin{figure}[!ht]
\centering
 \includegraphics[width=0.5\columnwidth]{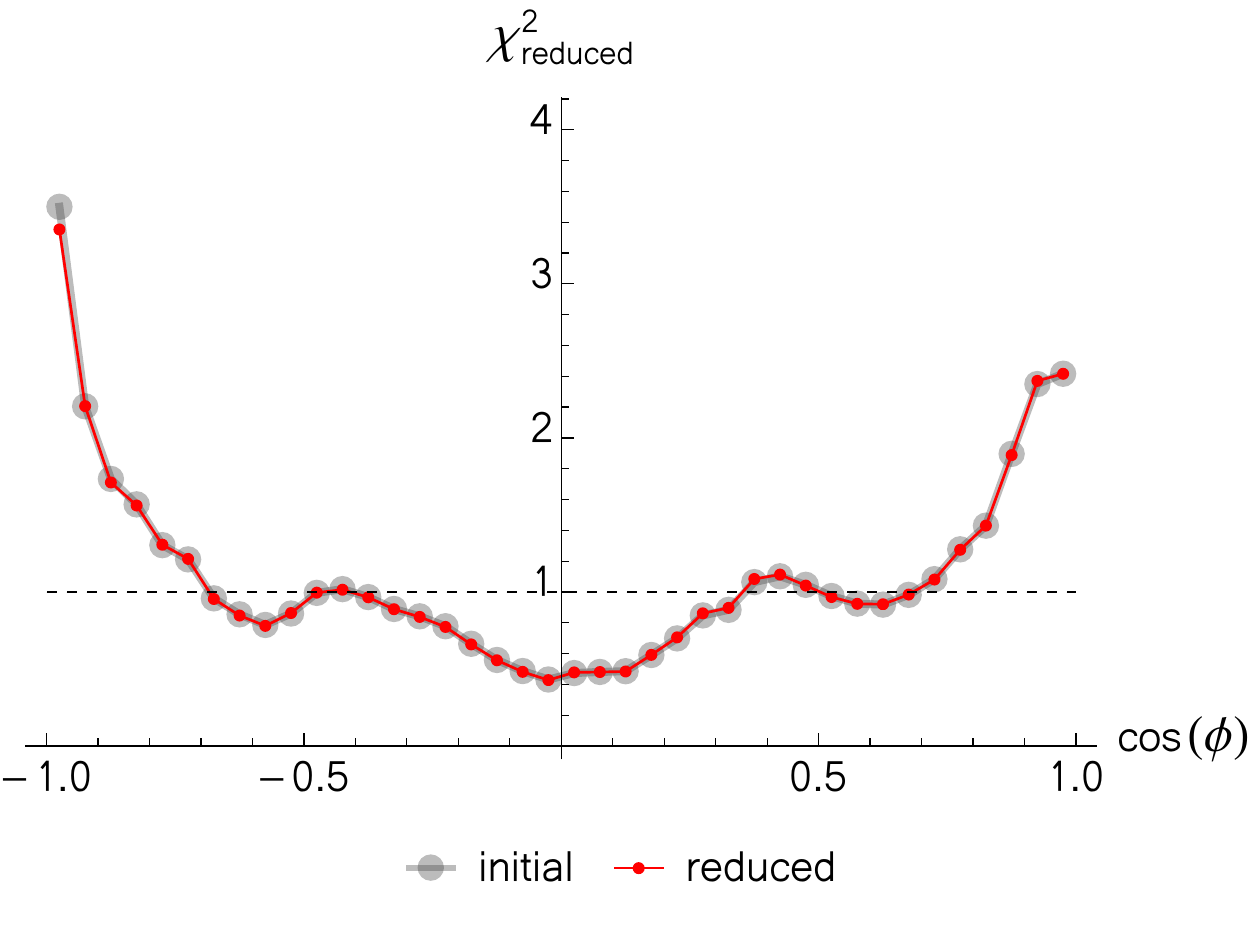}
	\caption{Comparison of reduced $\chi^2$ for the initial and reduced SW models as a function of $\cos(\phi)$. The reduced SW model is built on the first 9 PCs out of the 19 components obtained from the refitted SW model. Since only slight differences between the refitted and the original models are seen, it suggests that the order of the initial model is too high and that the model parameters are redundant. 
}
\label{fig:chi2}
\end{figure}

Now we check how elimination of the insignificant PCs impacts the model SW profiles by analyzing the reduced-model residuals as a function of heliographic co-latitude $\phi$ (i.e., $\phi = 0$ for the north pole and $\phi = \pi$ for the south pole). The residuals are arranged in 40 equi-areal latitudinal bins that correspond to bands on the solar sphere parallel to the solar equator. Selection of such bins reduces the impact of the inhomogeneous IPS data coverage near the polar regions, where the sparsity of the data and the relatively large spread of those that are available might lead to an overestimation of the residual dispersion. It was found that the profiles of the reduced SW model do not differ significantly from those in the refitted SW model, and that the reduction of the model dimensionality does not bias the results. This is illustrated in Figure \ref{fig:chi2}. It is seen that due to the PCA filtering the model accuracy is changed insignificantly.

Concluding this part of our research, we have re-formulated the refitted SW model obtained by \citet{porowski_etal:22a} and extended with data from 2021 into a reduced model, where all the yearly profiles are represented by the same number of 9 terms. The model is defined as a linear combination of the base functions obtained from the PCA analysis. This model can be used directly and supersedes that proposed by \citet{porowski_etal:22a}. In the further part of the paper, we focus on developing an approximation of this model using solar proxies.

\section{Proxy selection and processing}
\label{sec:proxySelection}
\noindent
Most of the solar proxy time series provide convenient measure of the solar activity evolution and show temporal variations that follow directly the solar cycle seasonality. On the other hand, the ACFs of the SW model parameters suggest that the parameters are characterized by a few independent types of variations. The fact that the SW parameters are deduced from high-resolution latitudinal profiles of the SW speed measured over many years suggests that other types of variabilities are of the
solar origin, since many aspects of the solar activity must have left their imprints on the SW profiles.
Therefore, we examine if the proxies also consist of components featuring different types of the solar seasonality, similar to those apparent in the SW model parameters. If so, they must be connected with the main variability and are less apparent in the proxy data because of their possible weak relative amplitude. Therefore, we expect that if we are able to separate other potential types of the intrinsic solar variations imprinted in the solar proxies, we will be able to link the proxies with the SW model parameters. Our objective in this subsection is to identify and disentangle the expected traces of variations in the proxies that are different than those related to the  11-year solar activity cycle, and to use them in extension of the SW model.

The proxies that may be potentially adopted for the model extension must fulfill the following prerequisites. The number of independent proxies must be similar to that of the base functions in the reduced SW model. At least some of the proxies must be correlated with the 11-year variation of the solar activity cycle. The proxy set must be able to reproduce the N/S asymmetry in the SW. The proxies must have a long time coverage, ideally extending before the beginning of the available IPS solar wind data (1985---2021), and be available nowadays. At least some of them must be connected with phenomena related to the solar magnetism, which is one of the factors responsible for the creation of the solar wind. 

\begin{figure}[!ht]
\center{
       \includegraphics[width=.95\columnwidth]{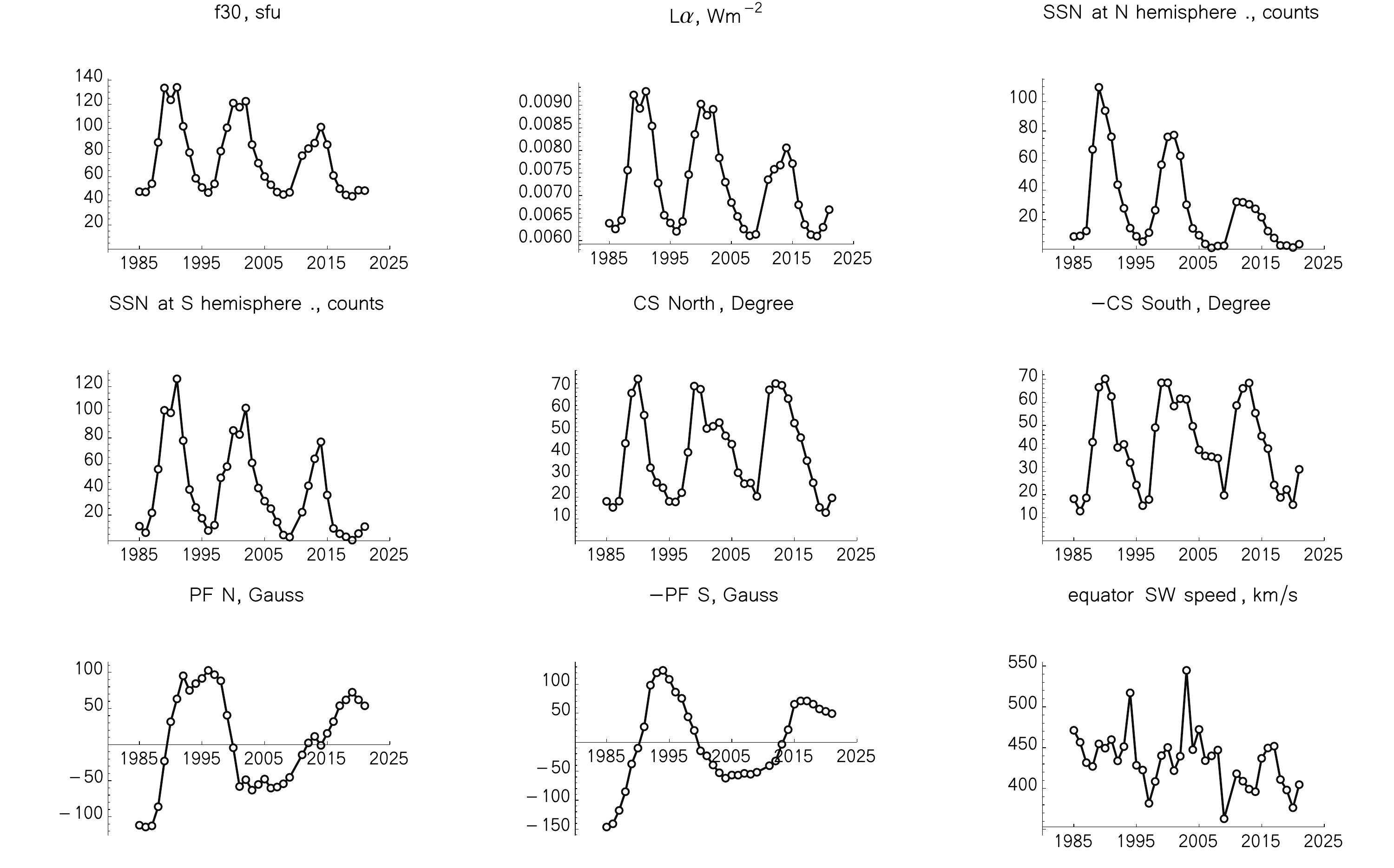}
    \caption{
	    Yearly averages of the proxies listed in Table \ref{tab:proxies} used in the final analysis.}
\label{fig:proxiesTS}
}
\end{figure}

On the other hand, it does not seem to be advantageous to admit several proxies very strongly correlated with each other, like, e.g., the F30 \citep{tanaka_kakinuma:57a, shimojo_etal:17a} and F10.7 \citep{tapping:13a} solar radio flux, because this would result in yielding little additional information to the system but would certainly add the inevitable noise to the system. \citet{fujiki_etal:15a} pointed out a usefulness of polar fields as proxies related to the solar wind structure.  We identified a proxy set that meets the requirements listed above and provides a satisfactory estimation of the SW profiles. The proxy set consists of the components listed in Table \ref{tab:proxies}, which also include proxies that had been considered as candidates during the tuning of the generalized model, but finally were rejected as a result of tests performed. Rejection of some of the candidate proxies was mostly because of their redundancy or negligible impact on the results. We found that the current sheet tilt angles in the north and south (CS$_{N/S}$) are essential for proper estimation of the SW speed profiles. Without adoption of the CS$_{N/S}$ tilts, the estimation of the profiles during the solar maxima shows a large discrepancy when compared with the profiles estimated with CS$_{N/S}$. At the moment, the CS$_{N/S}$ data availability limits the generalization of the SW model back to the year 1976 only. However, extension backward in time beyond 1976 will be possible if a proxy equivalent to CS$_{N/S}$ can be identified, for which data before 1976 are available. Also the strengths of the polar fields (PF$_{N/S}$) are important to obtain the proper results, which is a consequence of correlations between the polar fields and the fast/slow SW reported by \citet{tokumaru_etal:21a}. The time series of the selected proxy set is shown in Figure \ref{fig:proxiesTS}.

\begin{table}[!ht]
\caption{Basic properties of the used proxies.}
	\center
    \begin{tabular}{c|l|l|l|c|c|c}
	    \# in proxy setup & proxy & abbreviation & unit & data since & where measured & ref.\\
        \hline
	    1    &radio flux at 30 cm & F30 & sfu& 1951 & whole disk & [1]\\
     rejected    &radio flux at 10.7 cm & F107 & sfu& 1947 & whole disk & [2]\\
            2    &solar irradiance in Ly-$\alpha$ line  & L$\alpha$ & W m$^{-2}$& 1947   & whole disk & [3] \\
	    3,4    &sunspot number (N, S) & SSN$_{N/S}$ & counts &1755   & separate for N/S & [4]\\
        5,6 &current sheet tilt angle (N, S) & CS$_{N/S}$ & degree & 1976   & separate for N/S & [5]\\
     rejected    &MgII$_{c/w}$ ratio (Bremen) & MgII$_{c/w}$ & ratio & 1978   & whole disk & [6]\\
        7,8 &polar field strength (N, S) & PF$_{N/S}$  & G & 1976   &separate for N/S & [5]\\
	    9    &speed & SW speed & \kms &1963   & \it in situ\rm & [7]\\
      rejected   &density  & SW density & cm$^{-3}$ &1963   & \it in situ\rm & [7]\\
\end{tabular}
		\flushleft
     [1] \citet{tanaka_kakinuma:57a, shimojo_etal:17a,dudokdewit_etal:14a} 
		
		 [2] \citet{tapping:13a} 
		
		 [3] \citet{machol_etal:19a} 
		
		 [4] \citet{clette_etal:14a, veronig_etal:21b}
		
		 [5] \citet{hoeksema_etal:83a}
		
		 [6] \citet{viereck_etal:04a, snow_etal:14a}
		
		 [7] \citet{king_papitashvili:05}

        \label{tab:proxies}
\end{table}

Note that some of the selected proxies are correlated with the magnitude of the solar activity, and some of them with the strength of the solar magnetic field. While some of them are quantities averaged over the whole solar disk, some other ones are provided with discrimination between the north and south solar hemispheres. Additionally, proxies measured \it in situ \rm close to the solar equator are also used. These properties fulfill the requirements stated above.

The proxies were yearly-averaged to provide a temporal resolution identical to that of the IPS-based model, and standardized before use (i.e., the mean values were subtracted and the results divided by the respective standard deviations). Before averaging, the input proxy data at their full resolution were filtered against values derived from other proxies to remove potential artificial correlation enhancements.

\begin{figure}[!ht]
\center{
       \includegraphics[width=0.95\columnwidth]{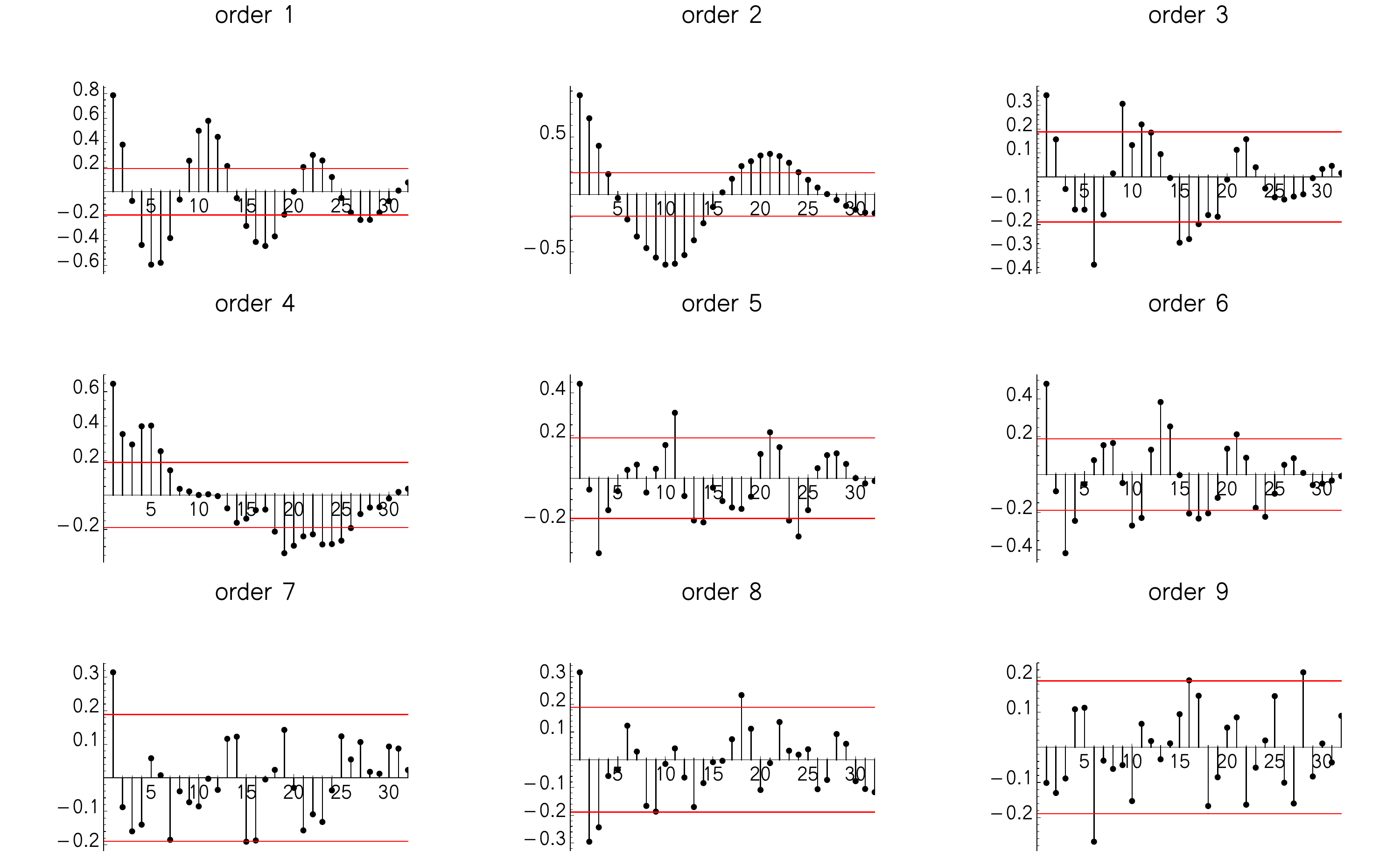}
    \caption{The ACF for the proxy PCs. The numbers above the panels indicate the order of the PC in the descending order of their contribution to the total variation.}
\label{fig:ccfProxies}
}
\end{figure}

Before application of the candidate proxies to the SW modeling, we attempted to identify and disentangle the common dominating variations existing in the proxy set into uncorrelated components and to eliminate potential redundancies that still might be present in the selected proxies. We processed the variations similarly as in the case of SW model parameters, i.e., we applied PCA to the full set of the selected proxies for the years 1985---2021. The ACF of the proxy PCA showed separation of the commonly known solar cycle periodicities in the first three PCs (Figure \ref{fig:ccfProxies}). In addition, the higher orders of the proxy PCs also showed evidence of weak seasonalities, but at the boundary of the adopted confidence level, also showing a moving average-like type of periodicity in the data set. Subsequently, we subjected the results of the proxy PCA to a dimension-reduction analysis, similar to that performed for the coefficients of the refitted SW model. However, the sensitivity of the predicted profiles provided by the generalization procedure to the proxy PC filtering is high. Therefore, the optimization of the proxy PC filtration level was performed with respect to the total accuracy of the generalization procedure, and simultaneously to the stability of the profile predictions provided by the generalization. We found that use of 6 of 9 proxy PCs in the fitting performed within the generalization protocol is optimal. The methodology of the optimization is discussed in section \ref{sec:tuning2}. 

\vfill
\newpage

\section{Results}
\label{sec:extendedModel}
\subsection{The extended SW model}
\noindent
Based on the development presented in the preceding sections, we propose an extended model of the evolution of the latitudinal structure of the SW, which is a base for the most recent version of the WawHelioIon 3DSW model. The extension is purely empirical, based on the model by \citet{porowski_etal:22a}, and has a yearly resolution in time and an infinite resolution in heliolatitude. Since the extension relies on fitting to all years at once, it comprises information about long-term evolution of the SW and offers an ability to predict SW speed profiles for times outside the IPS data interval. For the years for which IPS measurements are available, i.e., 1985---2021 (except 2010), it provides SW speed profiles in agreement with IPS observations, as shown with the red line in Figures \ref{fig:SWComp} and \ref{fig:SWCompB}. For the years outside this set, for which all necessary proxies are available, i.e., for the years 1976---1984, 
it provides proxy-based estimates of the 3D structure of the SW under assumption that no abrupt changes in the mean solar physical properties occurred during the year. 
The obtained proxy-based predictions are stable and show the SW speed evolution as expected qualitatively based on the phase of the solar activity cycle. The overall quality of the prediction ability of the proxy-based backward extension is confirmed by a comparison of the model predictions with the Ulysses \it in situ \rm measurements, as shown in the Figure \ref{fig:ulyssess}. 

The model extension is provided as a set of matrices and vectors. The set of matrices, according to Equation \ref{eq:YvsX}, includes the main matrix (\myvec{M}) and the vector of constants (\myvec{c}), shown in Equations \ref{eq:Mmatrix} and \ref{eq:cVector}, respectively. Additionally, a set of matrices and vectors necessary for preprocessing of the proxy data is provided (the matrix of proxy data must have the units and order as in Table \ref{tab:proxies}). This set  which includes:
(1) the normalization factors of the PCA ($\myvec{norm_{proxy}}$) and 
(2) matrix to calculate the proxy transformation to the frame of the PCA ($\myvec{M_{pca}}$),
shown, respectively, in Equations \ref{eq:norm} and \ref{eq:MpcaMatrix}.
As an example, a proxy matrix used in fitting (\myvec{x}) is shown below in Equation \ref{eq:proxyMatrix}.
The basis in the PCA frame and additional constants are in Equations \ref{eq:basis} and \ref{eq:basisConst}, respectively. 

\begin{equation}
\scriptsize
\myvec{M}=
\noindent
	\left(
\begin{array}{cccccc}
 48.8267 & 0.269098 & 30.5499 & 83.0085 & 23.4636 & -74.9528 \\
 -2.64821 & 1.55161 & 12.4453 & -7.55614 & -55.0053 & 25.0039 \\
 4.38350 & -9.04521 & -24.6587 & -14.8259 & -4.48943 & 27.1326 \\
 1.00285 & 2.12737 & -8.42190 & -4.81501 & 4.29399 & -44.6620 \\
 -0.180288 & -1.34206 & -10.4036 & 17.4988 & -2.80003 & -1.36492 \\
 0.995979 & 3.42453 & -2.50735 & -1.67740 & 18.7963 & 7.54459 \\
 0.697072 & -0.206675 & -11.1027 & 0.894392 & -5.52596 & 4.66782 \\
 -0.876336 & -2.11475 & -4.70134 & 6.53698 & -0.303955 & -27.0590 \\
 -1.12215 & 0.859762 & 1.55668 & 5.02860 & -1.91262 & -12.5170 \\
\end{array}
\right)
\normalsize
\label{eq:Mmatrix}
\end{equation}

\begin{equation}
\scriptsize
	\myvec{c}=\left(
\begin{array}{c}
 -416.545 \\
 -82.9764 \\
 209.221 \\
 159.309 \\
 129.691 \\
 -65.4379 \\
 209.373 \\
 79.4066 \\
 -27.9854 \\
\end{array}
\right)
\normalsize
\label{eq:cVector}
\end{equation}

\begin{equation}
\scriptsize
\myvec{norm_{proxy}}=\noindent\left(
\begin{array}{cc}
 75.2099 & 29.0239 \\
 0.00727234 & 0.00104463 \\
 28.3888 & 29.6817 \\
 39.9128 & 35.0266 \\
 40.6087 & 20.3532 \\
 41.5538 & 18.8522 \\
 4.88211 & 67.0727 \\
 2.14263 & 72.8142 \\
 434.605 & 35.3377 \\
\end{array}
\right)
\normalsize
\label{eq:norm}
\end{equation}

\begin{equation}
\scriptsize
\myvec{M_{pca}}=\noindent\left(
\begin{array}{cccccccccc}
 -0.421904 & -0.0733046 & -0.0429772 & 0.210449 & -0.104058 & 0.210305 & -0.118872 & -0.142205 & 0.825275 \\
 -0.421291 & -0.0710745 & -0.0201871 & 0.181684 & -0.184167 & 0.150249 & -0.139093 & -0.697529 & -0.470807 \\
 -0.398775 & -0.0187734 & -0.0502221 & 0.379336 & 0.763069 & -0.00682918 & 0.239255 & 0.187025 & -0.140237 \\
 -0.410258 & -0.0466386 & 0.00749083 & 0.311658 & -0.509035 & -0.257656 & -0.144500 & 0.584075 & -0.211657 \\
 -0.376287 & 0.00439128 & -0.112149 & -0.657751 & 0.239873 & 0.104671 & -0.553933 & 0.179313 & -0.0754078 \\
 -0.398439 & -0.00698783 & -0.0137800 & -0.493842 & -0.174061 & -0.176146 & 0.729091 & -0.0512587 & 0.0400257 \\
 0.0399996 & -0.691262 & 0.105116 & -0.0350202 & 0.132975 & -0.651507 & -0.150560 & -0.181070 & 0.103356 \\
 0.0533340 & -0.677785 & 0.216259 & -0.0497767 & -0.0675911 & 0.630974 & 0.151371 & 0.221111 & -0.118393 \\
 -0.111300 & 0.223033 & 0.961552 & -0.0381643 & 0.0614512 & -0.0467996 & -0.0663551 & -0.0255375 & 0.0284334 \\
\end{array}
\right)
\normalsize
\label{eq:MpcaMatrix}
\end{equation}

\begin{equation}
\scriptsize
\myvec{x}=\left(
\begin{array}{cccccccccc}
 47.5910 & 0.00638410 & 8.54607 & 11.2992 & 18.0214 & 18.1714 & -111.757 & -146.270 & 471.267 \\
 47.2946 & 0.00625642 & 9.01921 & 6.24068 & 15.2462 & 12.8308 & -114.194 & -140.750 & 456.647 \\
 54.2239 & 0.00645037 & 12.3194 & 21.7014 & 18.1154 & 18.5692 & -112.622 & -117.730 & 431.629 \\
 88.3978 & 0.00756160 & 67.4948 & 55.6484 & 44.6000 & 42.7500 & -86.1111 & -85.3889 & 427.035 \\
 133.484 & 0.00923105 & 109.550 & 101.526 & 67.6462 & 66.5769 & -22.8108 & -37.5135 & 454.652 \\
 123.582 & 0.00893128 & 93.7327 & 99.5513 & 74.1538 & 70.2231 & 31.6389 & -10.2222 & 449.350 \\
 133.984 & 0.00931042 & 76.0512 & 125.997 & 57.5286 & 62.5714 & 63.2703 & 26.5676 & 459.872 \\
 101.713 & 0.00854090 & 43.6122 & 77.9196 & 33.4385 & 40.4385 & 95.2778 & 98.3333 & 433.833 \\
 79.9851 & 0.00727459 & 27.6657 & 39.8245 & 26.6429 & 41.8071 & 75.0000 & 119.784 & 451.328 \\
 58.6667 & 0.00656255 & 14.2467 & 25.8597 & 24.2538 & 33.8923 & 84.5000 & 123.917 & 517.078 \\
 51.0805 & 0.00639097 & 8.70424 & 17.4646 & 17.9923 & 24.1308 & 91.6486 & 108.486 & 428.395 \\
 46.8648 & 0.00620282 & 5.09636 & 7.87204 & 17.7643 & 15.1929 & 102.778 & 86.4167 & 422.619 \\
 54.0940 & 0.00642495 & 11.1839 & 12.1327 & 22.0000 & 17.8769 & 96.8378 & 75.6486 & 381.851 \\
 81.1572 & 0.00746318 & 26.3806 & 48.9657 & 40.4929 & 49.0786 & 88.3889 & 43.7778 & 408.604 \\
 100.512 & 0.00835517 & 57.1578 & 57.7206 & 70.8462 & 68.5231 & 40.5676 & 20.2973 & 440.273 \\
 120.955 & 0.00902966 & 76.0164 & 85.8398 & 69.4231 & 68.5308 & -4.40541 & -14.8919 & 450.281 \\
 117.569 & 0.00878063 & 77.2718 & 82.6459 & 51.3857 & 58.3929 & -58.1111 & -23.4722 & 421.827 \\
 122.484 & 0.00891439 & 63.1818 & 103.289 & 52.4308 & 61.6308 & -48.6757 & -39.3243 & 439.493 \\
 86.5443 & 0.00783642 & 30.0755 & 60.6068 & 54.1286 & 61.3071 & -63.1389 & -52.7222 & 544.626 \\
 71.2132 & 0.00729561 & 14.0678 & 40.9891 & 48.0692 & 49.6769 & -55.2703 & -62.0541 & 447.536 \\
 60.1714 & 0.00684336 & 9.42059 & 30.8284 & 44.2231 & 39.4231 & -47.8056 & -56.6667 & 472.332 \\
 53.2975 & 0.00653530 & 3.49705 & 24.9660 & 31.1500 & 36.8000 & -60.2973 & -56.8378 & 434.105 \\
 47.2758 & 0.00625521 & 0.880995 & 14.5203 & 26.1077 & 36.4231 & -58.8333 & -53.7500 & 439.973 \\
 45.2347 & 0.00610638 & 2.14568 & 4.41786 & 26.4286 & 35.7786 & -54.3514 & -55.2973 & 447.116 \\
 47.0203 & 0.00613717 & 2.32493 & 2.78853 & 20.3077 & 19.6923 & -45.4167 & -51.8333 & 362.962 \\
 77.4265 & 0.00735210 & 32.0572 & 22.1852 & 69.1500 & 58.6857 & -14.2222 & -40.5278 & 418.096 \\
 83.3025 & 0.00758166 & 31.7112 & 42.7903 & 72.0154 & 66.0923 & 2.32432 & -32.3514 & 408.993 \\
 87.8305 & 0.00767330 & 30.3729 & 63.7661 & 71.2000 & 68.4462 & 11.3056 & -3.61111 & 399.240 \\
 101.088 & 0.00805843 & 27.2990 & 76.9838 & 65.0786 & 55.3786 & -1.10811 & 21.6216 & 396.081 \\
 86.5643 & 0.00770721 & 21.6115 & 35.5292 & 53.8846 & 45.3923 & 15.5000 & 65.5000 & 436.919 \\
 61.0023 & 0.00679223 & 12.2259 & 9.58289 & 47.2500 & 39.9571 & 31.8378 & 71.2432 & 449.602 \\
 50.0036 & 0.00635418 & 7.59104 & 5.42325 & 36.6615 & 24.2000 & 53.8611 & 71.2222 & 451.848 \\
 44.9223 & 0.00613083 & 2.51153 & 3.04444 & 26.5385 & 18.7462 & 61.8919 & 65.6757 & 410.959 \\
 43.7535 & 0.00609835 & 2.46505 & 0.421922 & 15.2500 & 22.1857 & 72.5000 & 57.3056 & 398.182 \\
 48.8119 & 0.00629570 & 1.14134 & 5.56315 & 12.8769 & 15.6000 & 62.0286 & 53.2286 & 376.457 \\
 48.4548 & 0.00668587 & 3.36790 & 10.9573 & 19.6125 & 30.9625 & 53.7297 & 49.3243 & 404.724 \\
\end{array}
	\right)
\normalsize
\label{eq:proxyMatrix}
\end{equation}

\begin{equation}
\resizebox{\textwidth}{!}{$
\myvec{basis}=
\left(
	\begin{array}{ccccccccccccccccccccc}
 0.0303615 & -0.148399 & -1.82016 & 0.950588 & -36.7423 & 6.26709 & 352.491 & -96.6696 & -1617.42 & 467.822 & 4459.57 & -1235.87 & -7855.94 & 1957.02
& 8964.88 & -1850.45 & -6438.14 & 963.036 & 2647.12 & -212.067 & -474.782 \\
 -0.154847 & 0.325799 & 3.90247 & 6.90727 & 21.6849 & -89.8171 & -280.281 & 287.864 & 1096.96 & 229.275 & -2508.66 & -3301.71 & 4024.62 & 8020.33
& -4760.62 & -9197.81 & 3870.65 & 5238.81 & -1848.41 & -1194.14 & 379.615 \\
 0.613839 & 0.372516 & 1.02147 & -33.3731 & 29.6560 & 630.252 & -747.087 & -4796.09 & 5842.06 & 19540.8 & -23826.1 & -46894.1 & 57197.7 & 68341.9
& -83658.0 & -59457.7 & 73319.5 & 28399.5 & -35397.0 & -5730.55 & 7237.97 \\
 0.145646 & -1.95972 & 10.6026 & 39.1579 & -189.615 & -562.665 & 1933.82 & 3921.91 & -11314.8 & -15399.3 & 39199.7 & 36557.9 & -83812.1 & -53479.0
& 112125. & 47073.4 & -91499.5 & -22837.6 & 41652.1 & 4687.62 & -8104.74 \\
 0.381407 & -1.57060 & -0.259320 & -26.3406 & -21.7187 & 728.965 & 355.245 & -5454.74 & -2696.69 & 20490.3 & 11624.3 & -44612.3 & -30199.8 & 58738.4
& 48036.0 & -46163.8 & -45673.7 & 19955.1 & 23785.1 & -3654.61 & -5208.56 \\
 0.0931676 & -0.754711 & -3.89398 & -31.1527 & 9.77272 & 396.029 & -91.4966 & -1877.23 & 1989.76 & 3758.42 & -12472.6 & -717.400 & 37180.8 & -9889.13
& -61203.7 & 17245.5 & 57173.6 & -11997.7 & -28442.7 & 3113.54 & 5859.90 \\
 0.843597 & 1.23274 & -3.23982 & 30.3795 & -79.4358 & -875.537 & 1012.62 & 7411.24 & -4466.28 & -31623.2 & 9048.70 & 78077.9 & -6323.86 & -116340.
& -7336.48 & 103177. & 17591.1 & -50126.2 & -12751.7 & 10266.8 & 3308.12 \\
 0.305436 & -0.662309 & 0.577370 & 25.6113 & 3.73271 & -277.483 & -298.114 & 1188.81 & 2384.84 & -1516.95 & -8910.03 & -4050.86 & 19456.6 & 16562.3
& -26484.2 & -23001.4 & 22236.6 & 14720.8 & -10563.3 & -3649.53 & 2172.33 \\
 0.00924623 & 0.975741 & -11.0824 & -43.9141 & 124.351 & 646.534 & -491.906 & -5116.36 & 701.214 & 23481.5 & 1051.08 & -63434.7 & -7397.01 & 101626.
& 17231.3 & -94738.9 & -21284.1 & 47447.5 & 13553.0 & -9868.64 & -3476.34 \\
 0.152456 & -0.188276 & -16.1914 & -24.2256 & 282.609 & 482.069 & -2465.05 & -3167.74 & 13203.7 & 10593.6 & -44965.7 & -21380.0 & 98114.0 & 27917.6
& -136403. & -23410.3 & 116763. & 11449.6 & -56029.1 & -2460.08 & 11515.3 \\
 0.542655 & -1.69619 & -37.5690 & 64.7035 & 828.075 & -748.126 & -7541.03 & 4261.83 & 35577.9 & -14448.9 & -96632.7 & 31480.6 & 157676. & -44762.6
& -153936. & 40136.4 & 84704.6 & -20506.6 & -22108.4 & 4523.67 & 1469.38 \\
 -0.0262717 & -2.90130 & -3.58938 & 67.2150 & 63.8465 & -510.549 & -590.883 & 690.347 & 4130.09 & 7796.28 & -20671.4 & -40722.7 & 64868.1 & 87368.0
& -121167. & -97371.6 & 130248. & 55470.6 & -74237.9 & -12784.5 & 17360.0 \\
 -0.232157 & -1.56151 & 28.5999 & 70.7166 & -692.226 & -1017.17 & 7124.31 & 7655.40 & -39606.8 & -33437.9 & 131569. & 88045.2 & -273097. & -141208.
& 357468. & 134653. & -286968. & -70039.4 & 129137. & 15279.8 & -24962.4 \\
 0.185056 & -0.209586 & -0.622268 & 37.5424 & -204.829 & -746.163 & 3389.47 & 6555.66 & -21340.2 & -30620.6 & 68362.2 & 82337.6 & -123165. & -131552.
& 127525. & 123287. & -72393.7 & -62578.4 & 19015.5 & 13278.9 & -1188.63 \\
 0.189349 & -0.689866 & -28.0463 & 18.4207 & 694.619 & -111.629 & -7229.82 & 28.5454 & 40173.9 & 1718.90 & -132734. & -6907.70 & 273528. & 13426.7
& -354699. & -14833.9 & 281029. & 8904.17 & -124143. & -2241.67 & 23407.5 \\
 0.0608202 & -1.19240 & -23.9073 & 15.6640 & 868.209 & 271.009 & -12072.2 & -4048.51 & 83475.3 & 20000.2 & -325806. & -49735.3 & 764414. & 68974.3
& -1.10008\times 10^6 & -53721.7 & 950520. & 21737.9 & -452518. & -3492.60 & 91219.5 \\
 -0.146085 & -7.06343 & 22.6386 & 325.455 & -780.398 & -4949.31 & 9955.51 & 36187.5 & -64163.4 & -147484. & 238589. & 358811. & -542647. & -532606.
& 766042. & 472488. & -654169. & -230004. & 309191. & 47237.9 & -62039.6 \\
 0.0422022 & 0.690534 & 7.56196 & -98.4124 & -179.249 & 2363.49 & 1388.67 & -22171.6 & -4741.18 & 106037. & 7283.10 & -289065. & -2659.44 & 467920.
& -5785.88 & -444809. & 6998.25 & 229203. & -2320.92 & -49379.7 & 9.42865 \\
 0.135358 & -2.76493 & -37.9291 & 168.453 & 1373.54 & -3056.17 & -18293.1 & 24551.3 & 121332. & -104910. & -457709. & 260684. & 1.04472\times 10^6
& -389442. & -1.46939\times 10^6 & 344887. & 1.24473\times 10^6 & -166870. & -582195. & 33991.5 & 115475. \\
\end{array}
\right)
\left(
\begin{array}{c}
 1 \\
 x \\
 x^2 \\
 x^3 \\
 x^4 \\
 x^5 \\
 x^6 \\
 x^7 \\
 x^8 \\
 x^9 \\
 x^{10} \\
 x^{11} \\
 x^{12} \\
 x^{13} \\
 x^{14} \\
 x^{15} \\
 x^{16} \\
 x^{17} \\
 x^{18} \\
 x^{19} \\
 x^{20} \\
\end{array}
\right)
$}
\label{eq:basis}
\end{equation}

\begin{equation}
	\scriptsize
	\myvec{basis_{const}}=\noindent\left(
\begin{array}{c}
 -1.39366 \\
 53.0259 \\
 -19.1181 \\
 1.79221 \\
 24.8439 \\
 -6.96403 \\
 -9.56386 \\
 -19.2455 \\
 24.4573 \\
 -6.01699 \\
\end{array}
\right)
\normalsize
	\label{eq:basisConst}
\end{equation}
\normalsize

Each row of \myvec{x} corresponds to a separate year and consists of yearly averages of individual proxies sorted in the ascending order (1985---2021, without 2010). The columns hold the time series of the individual proxies according to enumeration in Table \ref{tab:proxies}. Before use, the columns of \myvec{x} must be standardized using vector $\myvec{basis_{const}}$ (i.e., the arithmetic mean value must be subtracted from the time series in each column, and the results divided by the standard deviations of the time series).

In order to estimate a profile for the desired time using the matrices above, one must execute the following steps.

\noindent
(1) Obtain the proxy vector, with the components set according to the order and units listed in Table \ref{tab:proxies}. The mean time at which the proxy values were prepared will be the time for which the profile estimation is made.

\noindent
(2) Standardize the proxy vector using the mean and standard deviations stored in vector $\myvec{norm_{proxy}}$, in which the mean is the first column, and the standard deviation is the second column.

\noindent
(3) After standardization of the proxy vector, perform its PCA by calculating the following product: \\
$\myvec{proxyPC}=-(\myvec{proxyM}^{\text{T}}.\myvec{proxy}^{\text{T}})^{\text{T}}$, \\ 
and removing the last 3 values.

\noindent
(4) Calculate the following product, which returns the coefficients of the SW model: $\myvec{coeffs}=\myvec{c} + \myvec{M}.\myvec{modelProxyInput}$, and join the $\myvec{basis_{const}}$ to $\myvec{coeffs}$.

\noindent
(5) The profile is ready for insertion of a value from the range of $\cos(\phi)\in<-1,1>$. After calculating the following product: 
$profile(\cos(\phi))=(\myvec{coeffs}.\myvec{basis})(\phi)$,
one obtains a polynomial describing the mean SW profile. 

\subsection{Application}

\begin{figure}[!ht]
\center
       \includegraphics[width=.95\columnwidth]{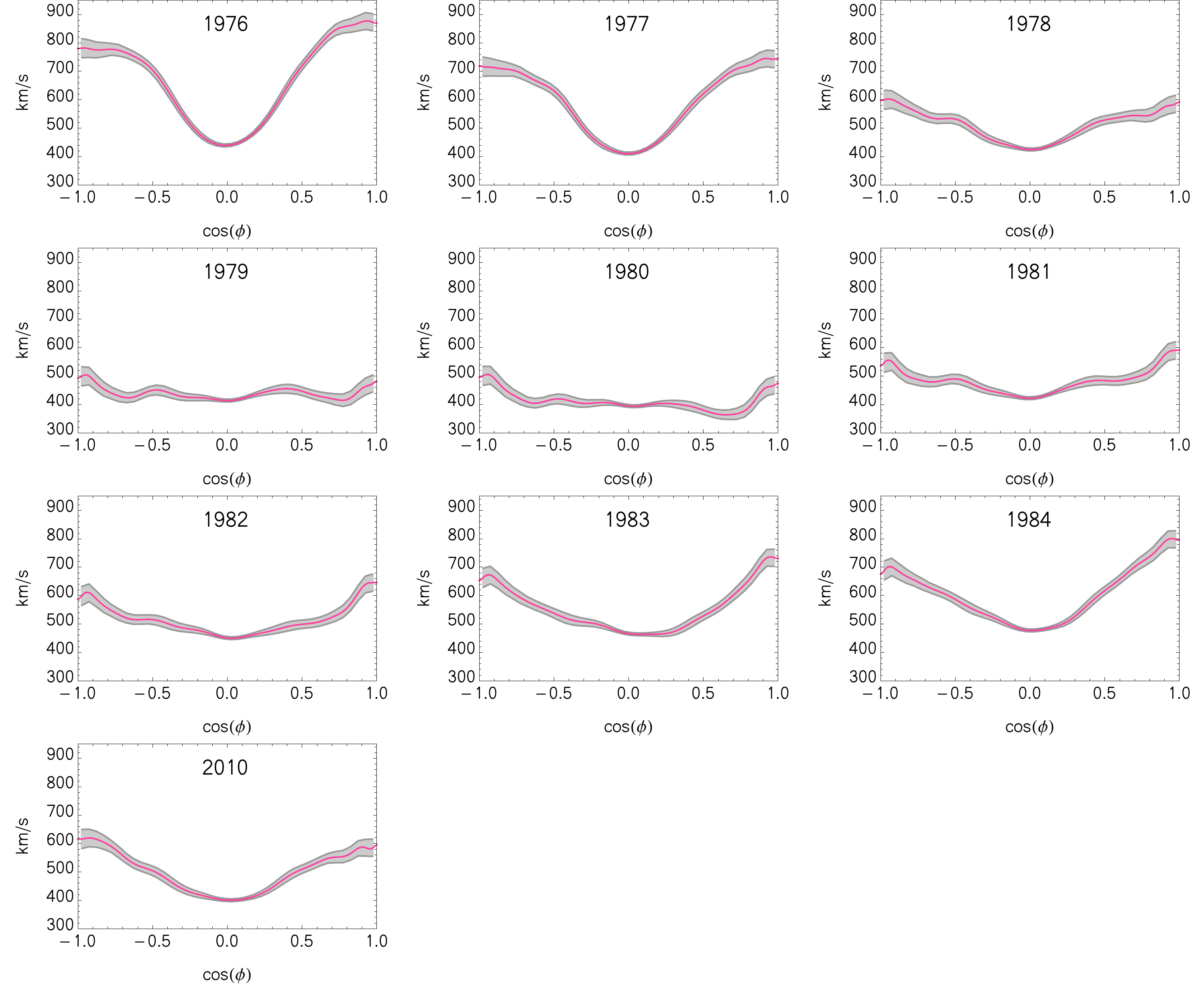}
	\caption{
		{Predictions of yearly mean profiles of the SW speed for the years 1976---1984 and 2010 obtained using the generalized SW speed model (red line). The shaded area indicates the 95\% confidence range.}
\label{fig:prediction}
}
\end{figure}

As an example of practical use of the derived correlation between the SW speed structure
and the solar proxies coded in the form of matrix \myvec{M},
we estimate yearly average 3D SW speed structures for years the 1976---1984 and 2010.
The above selection of years is made based on the availability of proxies,
and also to fill the IPS gap in 2010.
To obtain an estimation of the SW structure, we prepare yearly averages of the solar proxies, which we subsequently which normalize and calculate a product between the matrix \myvec{M} and the proxy vector, with an addition of the constant values that are also provided from the fit.
The overall description of practical application is described in the previous section.
The results of estimation are shown in Figure \ref{fig:prediction}.

The temporal variations of the predicted profiles seem exactly as expected, based on the solar activity history during the prediction years. The SW speed profile evolves from the shape characteristic for the solar minimum, which was about to end around 1976. During the following years, the profiles smoothly flatten at the maximum of Cycle 21, which occurred around 1980, to begin to rise at the poles, heralding a transition to another solar minimum, which occurred after 1985. This temporal behavior of the predictions for the years 1976---1984, based on the experience obtained from analysis of the solar cycles for which IPS data are available, suggests that they are correct. Also the prediction for year 2010, i.e., the year during which one expects flattening of the SW speed profile due to transition to the solar maximum epoch, was indeed obtained in these calculations.

\begin{figure}[!ht]
\center
       \includegraphics[width=0.49\columnwidth]{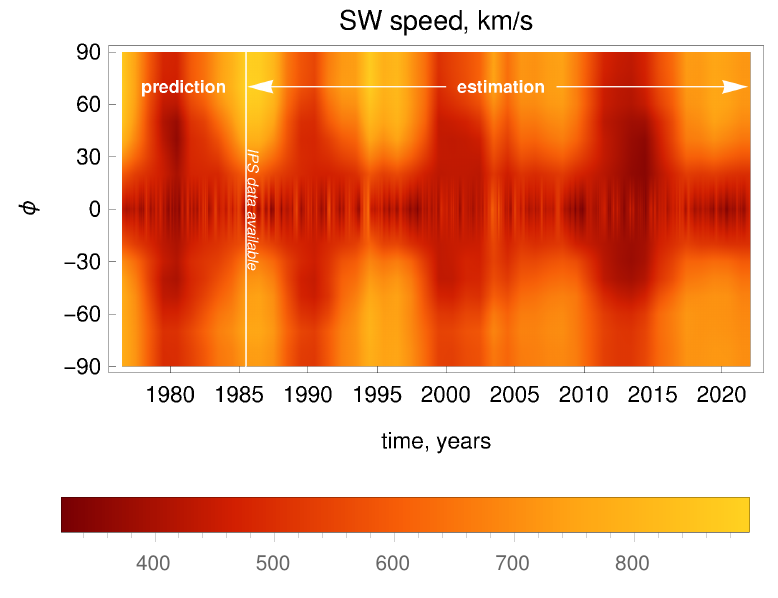}
       \includegraphics[width=0.49\columnwidth]{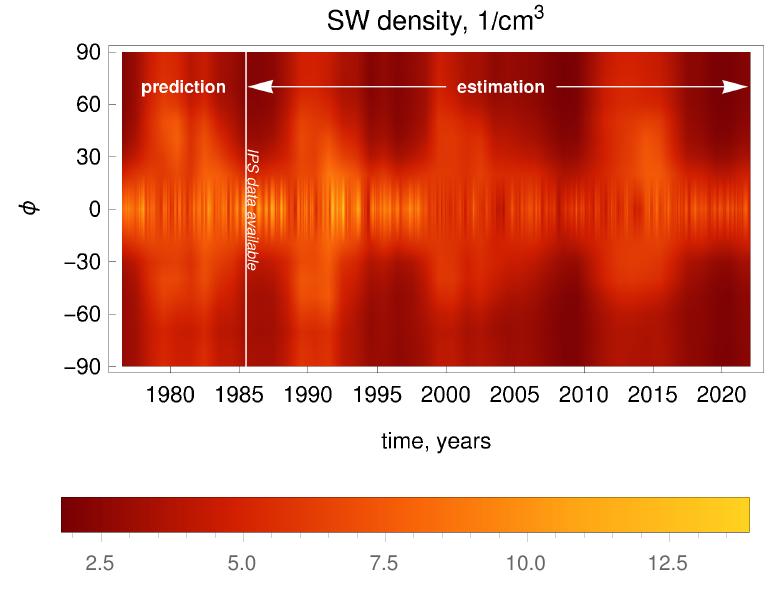}
\caption{Evolution of the SW speed and density profiles resulting from the model. The white vertical lines divide the maps into the area of predicted values (i.e., based on the proxies for the outside of the the IPS years 1976.5---1985.5) and the values obtained directly from the IPS analysis (1985.5---2022).
                }
\label{fig:maps}
\end{figure}

The calculation of the density is recommended to be performed identically as in \citep{porowski_etal:22a}, i.e., using the invariant SW energy flux estimated from the OMNI2 \it in situ \rm time series. Results of the speed and density calculation are presented in Figure \ref{fig:maps}.

\section{Discussion}
\label{sec:appliDiscu}
\noindent
The proxy-based model of the solar wind speed structure thus derived can be used to calculate yearly-averaged speed profiles of the solar wind for the times when IPS measurements of the SW speed are not available. Derivation calls for using yearly averages, but the time interval over which the averaging of the proxies will be performed may be selected according to specific needs. For example, the yearly profiles may be obtained not for the middle of a given calendar years, as we chose in this paper, but for any other part of a year. In particular, it can be centered at the halves of individual Carrington rotations. Thus, one might think of replacing the scheme used by \citet{porowski_etal:22a}, where the yearly-averaged SW parameter values obtained from IPS analysis are interpolated linearly for the centers of Carrington rotations, with a scheme where a speed profile for a given Carrington rotation is calculated from the proxies averaged over 13 Carrington rotations,
straddling this selected one.
 
This scheme cannot be expected to reproduce latitudinal profiles of the SW speed averaged over individual Carrington rotations. Analysis of longitude-averaged synoptic maps of the SW speed developed by CAT analysis \citep{tokumaru_etal:21a} suggests that changes of the SW speed profiles may be quite abrupt, occurring at time scale comparable to the Carrington period. Thus, while trends may be reproduced, most likely details will be missed.
 
An alternative approach could be calculating the speed profiles based on monthly averages of the proxies transformed using the PCA-derived matrix. A comparison of    the calculated profiles with IPS-derived data is outside the scope of this paper and will be performed in the future.
 
Potentially, the method derived in this paper may be applied to forecasting the solar wind structure. This could be performed by calculating the SW speed profiles based on the most recent proxies transformed with the PCA matrix. This approach might be useful, e.g., for calculating the survival probabilities of energetic neutral atoms \citep{bzowski:08a} observed by nowadays by IBEX, and in the future by IMAP mission \citep{mccomas_etal:18a} before the full set of IPS synoptic maps for a given year becomes available. While not definitive, this extrapolation could also facilitate analysis of the most recent ENA observations, at least tentatively.

\subsection{Impact of elimination of proxy PCs on the general accuracy of the model}
\label{sec:tuning2}
\noindent
In the matrix representation of the Generalized SW speed model (Eqn. \ref{eq:YvsX}),  matrix \myvec{M} contains the entire information on the multidimensional correlation between the SW model and solar proxies, obtained from fitting the input SW model to all IPS years: 1985---2021.
It was shown that the dimension of the input SW model may be reduced from 19 to 9 when the PC analysis is applied.
A similar dimension reduction procedure was also applied to the proxy PCs. 
It was motivated by an empirical study that showed that the least important proxy PCs do not improve the model but may contribute to propagation of uncertainties into the final solutions when projection of the model on years outside the IPS era is made, and also may lead to overfitting.
Here, we show the tuning of the filtering procedure of proxy PCs, aimed to find the optimal filtering level, at which the least important proxy PCs are eliminated from the fit,
but the remaining PCs preserve the salient information.
The tuning is a part of the overall iterative tuning procedure and is performed on the entire dataset.

\begin{figure}[!ht]
 \centering
  \includegraphics[width=0.45\columnwidth]{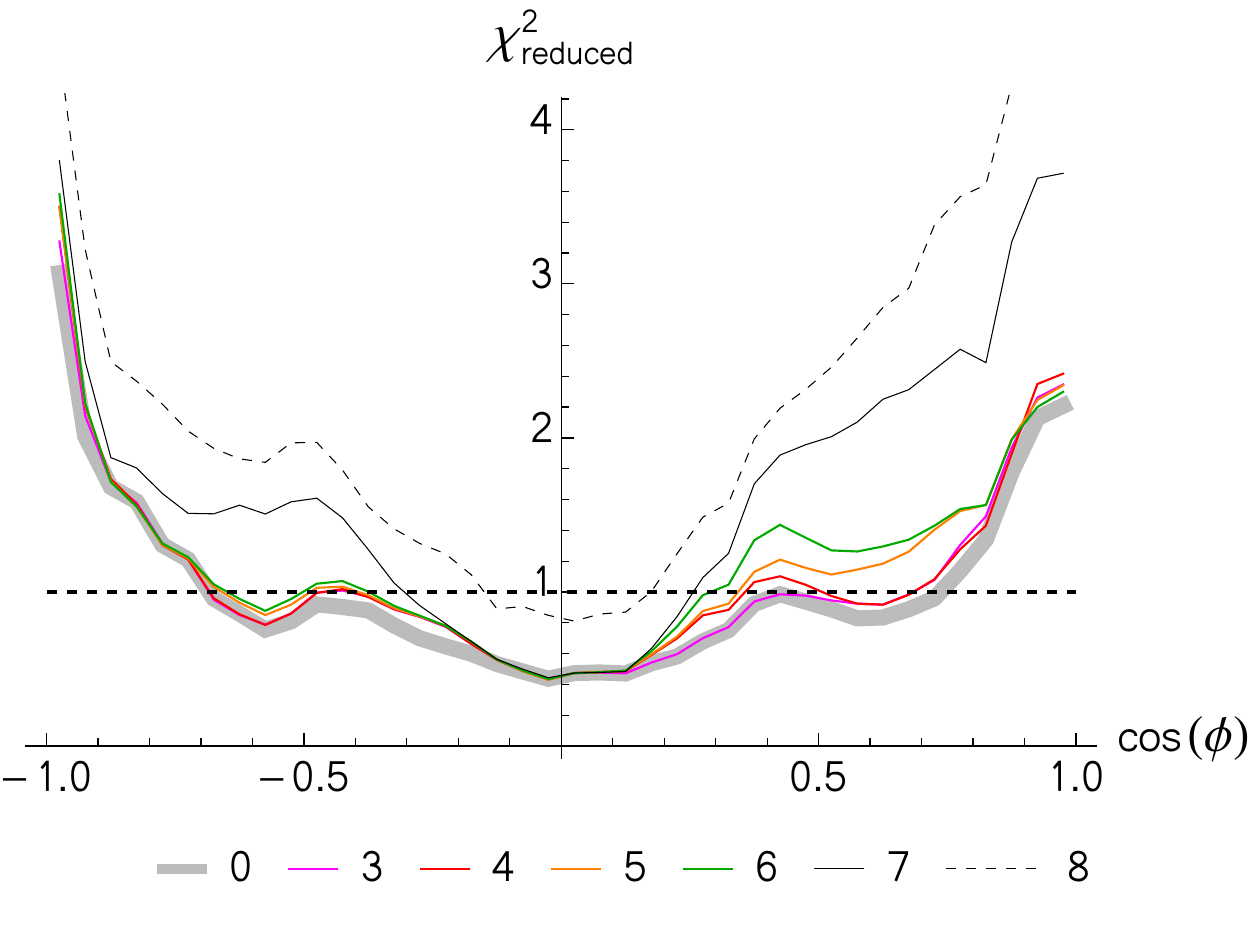}
     \caption{
         The magnitudes of $\chi^2$ of the SW speed profile reconstruction performed
         using the generalized SW model with respect to the
         IPS data for different levels of proxy PC's filtering. 
         Application of filtering levels above 3 causes deterioration of the fit quality.
         The filtering levels are marked below the plot; 0 means no filtering.
     }
 \label{fig:tuning2}
 \end{figure}

The impact of elimination of proxy PCs on the general accuracy is investigated using the $\chi^2$ measure.
We determine the $\chi^2$ magnitudes for different filtering levels of the proxy PCs used in the fits, starting the filtration from the less important proxy PC's (e.g., for the filtering level 3, the three least important proxy PC's are removed from the fit).
It was found that the fit quality starts to deteriorate above the filtration level 3. 
We thus presume that the filtering level 3 is optimum.
The results are summarized in the Figure \ref{fig:tuning2}.

\subsection{Analysis of the residuals}
\label{sec:internal}
\noindent
The deterioration of the fit quality towards the poles, which is seen in Figures \ref{fig:chi2} and \ref{fig:tuning2}, is analyzed in this section.
From our study we found that there are two most probable sources of the fit deterioration in the polar regions: worse IPS statistics in the polar regions or varying physical parameters of the solar wind.
It is commonly known that the polar regions have lower IPS statistics, which results in a reduced coverage of these regions on Carrington speed maps.
The use of equi-areal bins in heliolatitude during IPS binning before fitting reduces the impact of the lower IPS statistics on the general results, but the low statistics may still impact the fit quality in the case of fitting to all years at once.
On the other hand, empirical data show distinctive changes in the physical properties of the SW between the solar cycles observed in the IPS era. 
To qualify which of the two potential sources is more significant for the fit deterioration in the polar regions, we perform the following analysis.

We gathered the IPS data into three subsamples, which correspond to the full solar cycles during the IPS era.
We thus had the subsamples for SC22, SC23, and SC24.
For each of the three subsamples, we separately sought for correlation with the solar proxies (i.e., we determined three matrices $\myvec{M}_{\text{SC22}}$, $\myvec{M}_{\text{SC23}}$, $\myvec{M}_{\text{SC24}}$), and subsequently we reproduced the SW profiles for the years corresponding to a given matrix to compare the $\chi^2$ for all years with the $\chi^2$ for each of the three subsamples.

\begin{figure}[!ht]
 \centering
	\includegraphics[width=0.45\columnwidth]{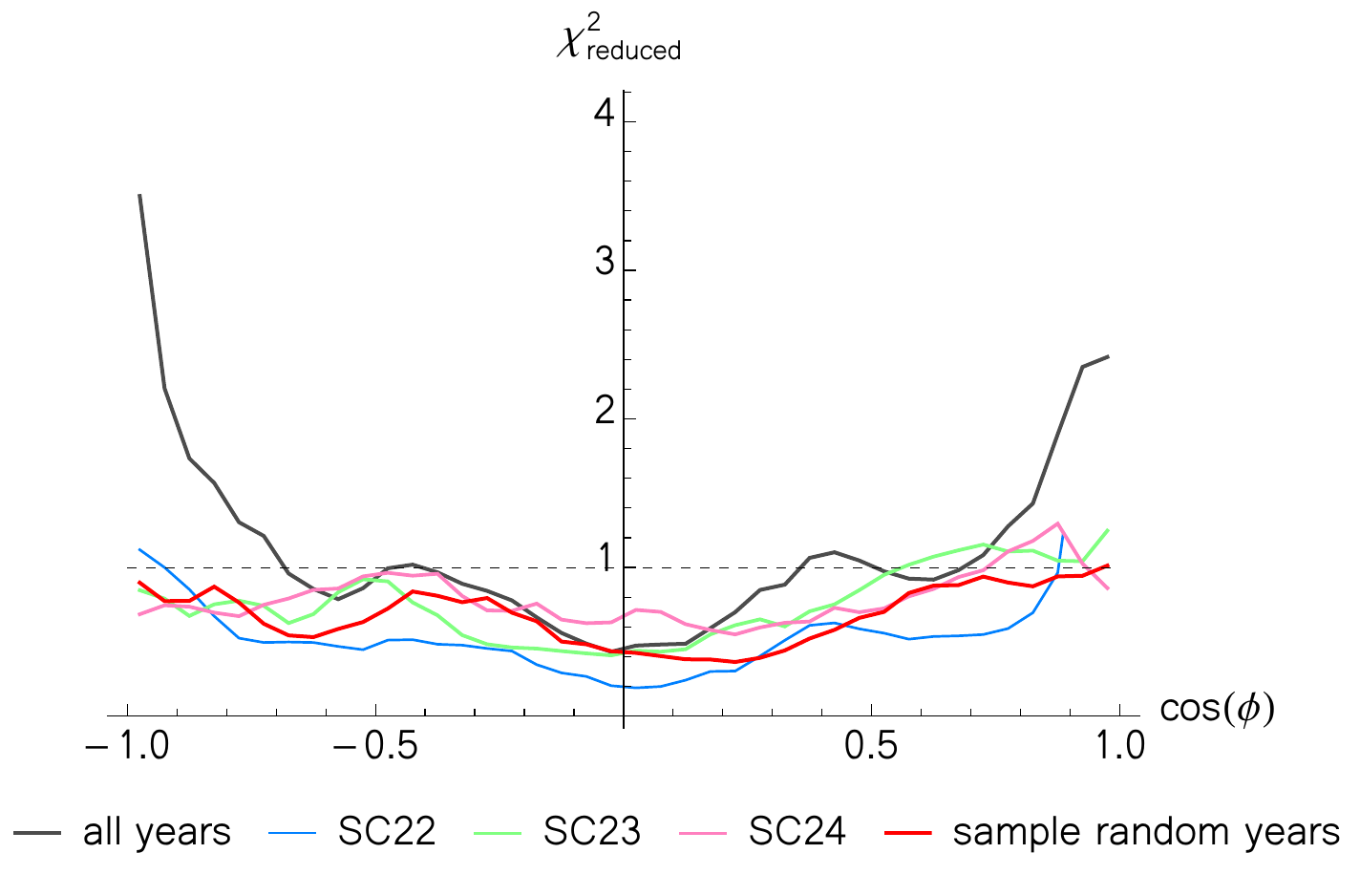}
     \caption{
         The magnitudes of $\chi^2$ of the SW speed profiles fitted to individual solar cycles (SC22, SC23, SC24), compared with a model fitted to a dataset of 11 randomly selected years.
     }
 \label{fig:chi2C}
 \end{figure}

It turned out that the deterioration on the poles is significantly lower for the full solar cycle subsamples of the IPS data.
To see if this is an effect of a lower statistics, we repeated the same operation and divided the entire sample into 11-year subsamples, but this time choosing the years randomly from the period 1985---2021, with each of the considered years selected only once.
We repeated this latter selection and fitting process 20 times. 
In both cases, i.e., when selecting the fit samples by solar cycles and randomly, we obtained a lower magnitude of $\chi^2$, i.e., a better quality of the fit around the polar regions (see Figure \ref{fig:chi2C}). 
This suggests that at this temporal resolution, we are not able to quantify to what extent the lower fit quality towards the poles is linked with the physical changes of the SW or with the reduced statistics.
It also indicates that the model based on the matrix \myvec{M} from the fit to all available IPS years should be adopted, i.e., no division of the IPS data into subsamples is recommended when a yearly-based approximation of the SW speed profile evolution is used in the fitting.

The last aspect of analysis of the residuals is the very low magnitude of reduced $\chi^2$ in the equatorial region. Both, the initial and filtered models have $\chi^2$ much below $1$. 
It is a result of application of the OMNI corrections on the IPS data in the situation when the OMNI SW speeds are subsequently used as one of the proxies. 
A contributing factor may be an overestimation of the data uncertainties due to the fact that their spread in the individual bins is not distributed exactly normally (i.e., Gaussian-like). This latter conclusion is drawn from analysis of Figure \ref{fig:chi2C}, where the magnitude of reduced $\chi^2$ is consistently below 1 for almost all data points.

\subsection{Comparison with the Ulysses data}
\label{sec:ulyssess}
\begin{figure}[!ht]
\centering
 \includegraphics[width=.9\columnwidth]{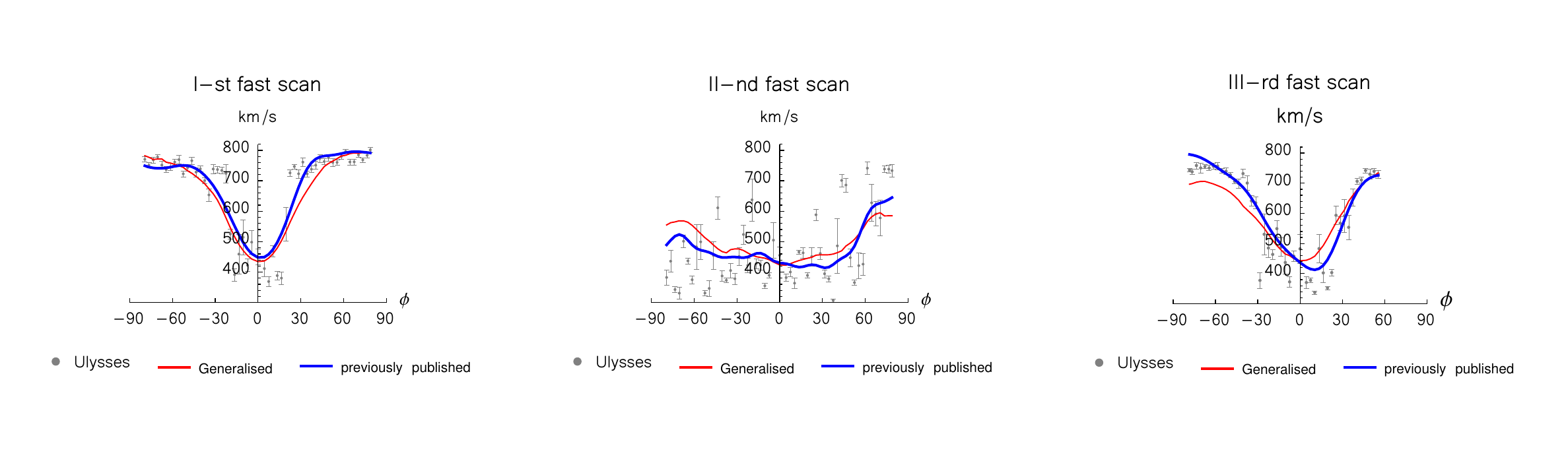}
    \caption{
	    Comparison of the SW speeds predicted using the generalized SW model (red), the SW speed model obtained in the previous work by \citet{porowski_etal:22a} (blue), and the averaged Ulysses data (black) for the three Ulysses fast latitude scans.}
\label{fig:ulyssess}
\end{figure}

\noindent
Now a comparison of model based on the matrix \myvec{M} its predictions with the Ulysses SW speeds measured \it in situ \rm \citep{mccomas_etal:98b, mccomas_etal:00b, mccomas_etal:02b, mccomas_etal:08a} is made. For the comparison purposes we use the three fast Ulysses scans and the yearly mean proxies as input to the generalized SW model. The results are shown in Figure \ref{fig:ulyssess}, in which a general agreement between the predictions and Ulysses data is seen. However, some minor differences occur, mostly at the areas with large dynamic changes of SW. The problems with reproduction of the high dynamics of the SW speed changes is an issue of the generalization procedure, which was pointed out in Section \ref{sec:internal}. This issue is due to the fact that the generalization procedure relies on yearly mean profiles, and and the fitting is done for all years at once. It causes that the generalization may sometimes have a large inertia, which results in a worsening of the accuracy of the predictions because of a slow adaptation of the model to the rapidly changing SW. Nevertheless, the differences in the first and second fast scan are mostly within the uncertainties of the Ulysses data.

The third fast scan, which was performed in 2007, features some systematic differences in the southern hemisphere. In general, the solar wind in 2007 was quite unique, since the SW speed profile featured a large N/S asymmetry close to the equator, which is also confirmed by the IPS data (see Figure \ref{fig:SWCompB}, where the asymmetry in the original SW reconstruction in the vicinity of the equator by \citet{porowski_etal:22a} is also visible). Since such a situation does not appear frequently during the other years, the generalization procedure was not able to reproduce this asymmetry and provided a systematic difference for this specific year. Likely, this difference might be reduced if data for a longer time interval had been available or a larger temporal resolution would be used during the fitting performed within the generalization procedure, assuming that the unusual asymmetric of profile was not due to some other factors.

Summarizing, the generalization procedure will always be insensitive to rare events of yearly SW speed profile shape distortions, and only a larger input data with numerous occurrences of the rare SW long-scale configurations will make this procedure sensitive to them. Despite these drawbacks, the profits offered by generalization  and the fact that unusual profiles are rather rare, make it a useful tool for analyses of the SW speed.

\section{Summary and conclusions}
\label{sec:Summary}
\noindent
We discovered an empirical correlation between the profiles of SW speed and a set of solar proxies, which allowed us to extend the model of evolution of the latitudinal structure of the SW developed by \citet{porowski_etal:22a} into times for which the IPS measurements are not available. We called this extension a Generalized SW Model. We showed that the Generalized SW Model provides relatively stable solutions, even among different model setups. The solutions provided by the Generalized Model converge when PCA filtering is applied to the SW model parameters (with the 1\% criterion) and to the proxies. The proxy PCA filtering was carefully optimized based on analysis of the convergence of the solutions. This holds for a wide range of $m$ values (8---22), as well as for small variations of the proxy set composition.
The overall accuracy of the generalization is confirmed by comparing its predictions with the Ulysses data, as shown in Figure \ref{fig:ulyssess}, which demonstrates the fidelity of the model.

The model calculates the profiles of the SW speed for the middle of a calendar year. For the years for which IPS measurements are available, the profiles are derived based on the IPS data; for those for which they are not available, the proxy-based model is used. The evolution of the speed and density of the SW resulting from the model for the entire time interval for which the IPS or proxy data are available, i.e., for 1976---2022, are presented in Figure \ref{fig:maps}.

\noindent The analysis steps leading to the final model are summarized as follows:
\begin{enumerate}
    \item Take synoptic Carrington maps of the solar wind speed retrieved from CAT analysis performed by \citet{tokumaru_etal:21a}on IPS observations. Perform IPS data preparation as presented by \citet{porowski_etal:22a}. Calculate Carrington-averaged profiles of the solar wind speed cleaned from outliers, averaged over uniform latitudinal bands in the space of $\cos(\phi)$, where $\phi$ is heliolatitude.

\item Define the Legendre base in the $\cos \phi$ space with the constraint of disappearing the derivatives over $\phi$ at the poles, as defined in Equations 1, 2 in \citet{porowski_etal:22a}, for the order $n = 20$ and fit the parameters of the Legendre model for all of the available $m$ years of the IPS data. The result is an $m$ by $n-1$ matrix of the model parameters.

\item Perform the PCA on this matrix. Take the resulting matrix of the transformation from the parameter space to the PC space and transform the Legendre base into the PCA base. Reduce the dimension of this base by admitting only the base vectors corresponding to the PCs that contribute at least 1\% to the total variance (in our case, the 9 largest PCs); adopt these base vectors and the parameters as the basis for the reduced SW model. 

\item Take the 9 proxies listed in Table \ref{tab:proxies}, calculate yearly averages for all of them for the years that have the IPS measurements. The result is a matrix of $m$ by 9 yearly averages of the proxy values.

\item Perform the PCA on the proxies, transform the original proxy time series matrix into the PCA space.

\item Perform fitting of the speed profiles for the proxies from the time intervals when there are no IPS data available; to that end, use Equation \ref{eq:YvsX} to obtain $\myvec{M}$ and $\myvec{c}$. The result is SW speed profiles for the centers of calendar years.

\item Calculate the SW density profiles using the just-derived speed profiles based on the total SW energy flux \citep{leChat_etal:12a} using the methodology employed  by \citet{porowski_etal:22a}.
\end{enumerate}

We checked that the Generalized SW Model provides stable solutions both within the IPS data range (1985---2021) and outside this period. We found no systematic differences, but noticed an increasing spread of the residuals towards the solar poles. We found that this spread is neither due to the adopted model setup, nor to the proxy set used. We concluded that the Generalized SW Model for the 36 years of available IPS data returns some discrepancies at higher latitudes, especially during periods of high dynamics of the SW variations (i.e., during transition between the minima and maxima of the solar activity). Most likely, a larger temporal resolution of the initial SW model would be necessary to eliminate this effect. 

On the other hand, the global accuracy of the SW profiles generated by the Generalized SW Model is satisfactory. The Generalized SW Model may be regarded as a useful tool to provide SW speed distribution for periods when no IPS observations are available.
At the current state of analysis it seems impossible to conclude if an improvement of the $\chi^2$ of the model observed when 11-year subsamples are used in the fitting is a statistical effect or if this is because the the relation between the SW parameters and the proxies used in the Generalized Model changes from one solar cycle to another. A resolution of this enigma may be obtained when the time resolution of the Generalized model is increased, which will be a topic of a separate
study.

The Generalized SW Model is the newest extension of the WawHelIioIon 3DSW model of the evolution of the solar wind speed and density, and the resulting charge exchange rates of interstellar neutral species inside the heliosphere. This extension provides an opportunity to improve the current knowledge on the solar wind and its interaction with the interstellar medium.

\begin{acknowledgments}
Acknowledgments \\
The IPS observations were conducted under the solar wind program of the Institute for Space-Earth Environmental Research (ISEE) of Nagoya University.
The OMNI data were obtained from the GSFC / SPDF OMNIWeb interface at https://omniweb.gsfc.nasa.gov. The F30 and F10.7 radio fluxes were obtained  online (https://spaceweather.cls.fr/services/radioflux/).
The Nobeyama Radio Polarimeters (NoRP) are operated by Solar Science Observatory, a branch of National Astronomical Observatory of Japan, and their observing data are verified scientifically by the consortium for NoRP scientific operations.
The Version 4 of solar Ly-$\alpha$ composite (https://doi.org/10.25980/zr1t-6y72) is available from the University of Colorado Laboratory for Atmospheric and Space Physics (LASP) Interactive Solar Irradiance Datacenter (LISIRD; http://lasp.colorado.edu/lisird/data).
The sunspot data were taken from the World Data Center SILSO, Royal Observatory of Belgium, Brussels.
Wilcox Solar Observatory CS$_{N/S}$ and PF$_{N/S}$ data used in this study were obtained via the web site http://wso.stanford.edu.
The Wilcox Solar Observatory is currently supported by NASA.
The University of Bremen Mg II Index is available online at http://www.iup.uni-bremen.de/UVSAT/Datasets/mgii or via LISIRD database.
This study was supported by Polish National Science Center grant 2019/35/B/ST9/01241.
\end{acknowledgments}

\bibliography{iplbib}{}

\begin{thebibliography}{}
\expandafter\ifx\csname natexlab\endcsname\relax\def\natexlab#1{#1}\fi
\providecommand{\url}[1]{\href{#1}{#1}}
\providecommand{\dodoi}[1]{doi:~\href{http://doi.org/#1}{\nolinkurl{#1}}}
\providecommand{\doeprint}[1]{\href{http://ascl.net/#1}{\nolinkurl{http://ascl.net/#1}}}
\providecommand{\doarXiv}[1]{\href{https://arxiv.org/abs/#1}{\nolinkurl{https://arxiv.org/abs/#1}}}

\bibitem[{{Bame} {et~al.}(1992){Bame}, {McComas}, {Barraclough}, {Phillips},
  {Sofaly}, {Chavez}, {Goldstein}, \& {Sakurai}}]{bame_etal:92a}
{Bame}, S.~J., {McComas}, D.~J., {Barraclough}, B.~L., {et~al.} 1992, \aaps,
  92, 237

\bibitem[{Bzowski(2003)}]{bzowski:03}
Bzowski, M. 2003, \aap, 408, 1155, \dodoi{10.1051/0004-6361:20031023}

\bibitem[{Bzowski(2008)}]{bzowski:08a}
---. 2008, \aap, 488, 1057, \dodoi{10.1051/0004-6361:200809393}

\bibitem[{Bzowski \& Kubiak(2020)}]{bzowski_kubiak:20a}
Bzowski, M., \& Kubiak, M.~A. 2020, \apj, 901, \dodoi{10.3847/1538-4357/abada2}

\bibitem[{Bzowski {et~al.}(2003)Bzowski, M{\"a}kinen, Kyr{\"o}l{\"a}, Summanen,
  \& Qu{\`e}merais}]{bzowski_etal:03a}
Bzowski, M., M{\"a}kinen, T., Kyr{\"o}l{\"a}, E., Summanen, T., \&
  Qu{\`e}merais, E. 2003, \aap, 408, 1165, \dodoi{10.1051/0004-6361:20031022}

\bibitem[{Bzowski {et~al.}(2002)Bzowski, Summanen, Ruci{\'n}ski, \&
  Kyr{\"o}l{\"a}}]{bzowski_etal:02}
Bzowski, M., Summanen, T., Ruci{\'n}ski, D., \& Kyr{\"o}l{\"a}, E. 2002, \jgr,
  107, 1101, \dodoi{10.1029/2001JA000141}

\bibitem[{Bzowski {et~al.}(2013)Bzowski, Sok{\'{o}}{\l}, Tokumaru, Fujiki,
  Qu{\'e}merais, Lallement, Ferron, Bochsler, \& McComas}]{bzowski_etal:13a}
Bzowski, M., Sok{\'{o}}{\l}, J.~M., Tokumaru, M., {et~al.} 2013, in
  {Cross-Calibration of Far {UV} Spectra of Solar Objects and the Heliosphere},
  ed. E.~Qu{\'e}merais, M.~Snow, \& R.~Bonnet, {ISSI Scientific Report} No.~13
  ({Springer Science+Business Media}), 67--138, doi
  10.1007/978--1--4614--6384--9$\_$3, \dodoi{10.1007/978-1-4614-6384-9_3}

\bibitem[{{Chalov} {et~al.}(2003){Chalov}, {Fahr}, \&
  {Izmodenov}}]{chalov_etal:03a}
{Chalov}, S.~V., {Fahr}, H.~J., \& {Izmodenov}, V.~V. 2003, \jgr, 108, 1266,
  \dodoi{10.1029/2002JA009492}

\bibitem[{Clette {et~al.}(2014)Clette, Svalgaard, Vaquero, \&
  Cliver}]{clette_etal:14a}
Clette, F., Svalgaard, L., Vaquero, J., \& Cliver, E. 2014, \ssr, 186, 35,
  \dodoi{10.1007/s11214-014-0074-2}

\bibitem[{{Dayeh} {et~al.}(2019){Dayeh}, {Zirnstein}, {Desai}, {Funsten},
  {Fuselier}, {Heerikhuisen}, {McComas}, {Schwadron}, \&
  {Szalay}}]{dayeh_etal:19a}
{Dayeh}, M.~A., {Zirnstein}, E.~J., {Desai}, M.~I., {et~al.} 2019, \apj, 879,
  84, \dodoi{10.3847/1538-4357/ab21c1}

\bibitem[{{Dudok de Wit} {et~al.}(2014){Dudok de Wit}, {Bruinsma}, \&
  {Shibasaki}}]{dudokdewit_etal:14a}
{Dudok de Wit}, T., {Bruinsma}, S., \& {Shibasaki}, K. 2014, Journal of Space
  Weather and Space Climate, 4, A260000, \dodoi{10.1051/swsc/2014003}

\bibitem[{{Fahr} {et~al.}(2016){Fahr}, {Sylla}, {Fichtner}, \&
  {Scherer}}]{fahr_etal:16a}
{Fahr}, H.-J., {Sylla}, A., {Fichtner}, H., \& {Scherer}, K. 2016, \jgr, 121,
  8203

\bibitem[{{Florinski} {et~al.}(2016){Florinski}, {Heerikhuisen}, {Niemiec}, \&
  {Ernst}}]{florinski_etal:16a}
{Florinski}, V., {Heerikhuisen}, J., {Niemiec}, J., \& {Ernst}, A. 2016, \apj,
  826, 197, \dodoi{10.3847/0004-637X/826/2/197}

\bibitem[{{Fujiki} {et~al.}(2015){Fujiki}, {Tokumaru}, {Iju}, {Hakamada}, \&
  {Kojima}}]{fujiki_etal:15a}
{Fujiki}, K., {Tokumaru}, M., {Iju}, T., {Hakamada}, K., \& {Kojima}, M. 2015,
  \solphys, 290, 2491, \dodoi{10.1007/s11207-015-0742-8}

\bibitem[{{Galli} {et~al.}(2019){Galli}, {Wurz}, {Rahmanifard}, {M{\"o}bius},
  {Schwadron}, {Kucharek}, {Heirtzler}, {Fairchild}, {Bzowski}, {Kubiak},
  {Kowalska-Leszczy{\'n}ska}, {Sok{\'o}{\l}}, {Fuselier}, {Swaczyna}, \&
  {McComas}}]{galli_etal:19a}
{Galli}, A., {Wurz}, P., {Rahmanifard}, F., {et~al.} 2019, \apj, 871, 52,
  \dodoi{10.3847/1538-4357/aaf737}

\bibitem[{{Giacalone} {et~al.}(2021){Giacalone}, {Nakanotani}, {Zank},
  {K{\`o}ta}, {Opher}, \& {Richardson}}]{giacalone_etal:21a}
{Giacalone}, J., {Nakanotani}, M., {Zank}, G.~P., {et~al.} 2021, \apj, 911, 27,
  \dodoi{10.3847/1538-4357/abe93a}

\bibitem[{{Gloeckler} {et~al.}(1992){Gloeckler}, {Geiss}, {Balsiger}, {Bedini},
  {Cain}, {Fisher}, {Fisk}, {Galvin}, {Gliem}, \&
  {Hamilton}}]{gloeckler_etal:92}
{Gloeckler}, G., {Geiss}, J., {Balsiger}, H., {et~al.} 1992, \aaps, 92, 267

\bibitem[{{Gruntman} {et~al.}(2001){Gruntman}, {Roelof}, {Mitchell}, {Fahr},
  {Funsten}, \& {McComas}}]{gruntman_etal:01a}
{Gruntman}, M., {Roelof}, E.~C., {Mitchell}, D.~G., {et~al.} 2001, \jgr, 106,
  15767, \dodoi{10.1029/2000JA000328}

\bibitem[{{Gruntman}(1997)}]{gruntman:97}
{Gruntman}, M.~A. 1997, Review of Scientific Instruments, 68, 3617,
  \dodoi{10.1063/1.1148389}

\bibitem[{{Heerikhuisen} {et~al.}(2010){Heerikhuisen}, {Pogorelov}, {Zank},
  {Crew}, {Frisch}, {Funsten}, {Janzen}, {McComas}, {Reisenfeld}, \&
  {Schwadron}}]{heerikhuisen_etal:10a}
{Heerikhuisen}, J., {Pogorelov}, N.~V., {Zank}, G.~P., {et~al.} 2010, \apjl,
  708, L126, \dodoi{10.1088/2041-8205/708/2/L126}

\bibitem[{{Hewish} {et~al.}(1964){Hewish}, {Scott}, \&
  {Wills}}]{hewish_etal:64a}
{Hewish}, A., {Scott}, P.~F., \& {Wills}, D. 1964, \nat, 203, 1214,
  \dodoi{10.1038/2031214a0}

\bibitem[{Hoeksema {et~al.}(1983)Hoeksema, Wilcox, \&
  Scherrer}]{hoeksema_etal:83a}
Hoeksema, J.~T., Wilcox, J.~M., \& Scherrer, P.~H. 1983, \jgr, 88, 10331,
  \dodoi{10.1029/JA088iA12p09910}

\bibitem[{Izmodenov \& Alexashov(2015)}]{izmodenov_alexashov:15a}
Izmodenov, V.~V., \& Alexashov, D.~B. 2015, \apjs, 220, 32,
  \dodoi{10.1088/0067-0049/220/2/32}

\bibitem[{{Jackson} {et~al.}(1997){Jackson}, {Hick}, {Kojima}, \&
  {Yokobe}}]{jackson_etal:97a}
{Jackson}, B.~V., {Hick}, P.~L., {Kojima}, M., \& {Yokobe}, A. 1997, \asr, 20,
  23, \dodoi{10.1016/S0273-1177(97)00474-2}

\bibitem[{Jackson {et~al.}(1998)Jackson, Hick, Kojima, \&
  Yokobe}]{jackson_etal:98a}
Jackson, B.~V., Hick, P.~L., Kojima, M., \& Yokobe, A. 1998, \jgr, 103, 12049,
  \dodoi{10.1029/97JA02528}

\bibitem[{Jolliffe(2004)}]{jolliffe:04a}
Jolliffe, I.~T. 2004, {Principal Component Analysis}, second edition (New York,
  Berlin, Heidelberg: Springer)

\bibitem[{Katsavarias {et~al.}(2012)Katsavarias, Preka-Papadema, \&
  Moussas}]{katsavarias_etal:12a}
Katsavarias, C., Preka-Papadema, P., \& Moussas, X. 2012, \solphys, 280, 623,
  \dodoi{10.1007/s11207-012-0078-6}

\bibitem[{{King} \& {Papitashvili}(2005)}]{king_papitashvili:05}
{King}, J.~H., \& {Papitashvili}, N.~E. 2005, \jgr, 110, 2104,
  \dodoi{10.1029/2004JA010649}

\bibitem[{{Kojima} \& {Kakinuma}(1990)}]{kojima_kakinuma:90a}
{Kojima}, M., \& {Kakinuma}, T. 1990, \ssr, 53, 173, \dodoi{10.1007/BF00212754}

\bibitem[{Kojima {et~al.}(1998)Kojima, Tokumaru, Watanabe, Yokobe, Asai,
  Jackson, \& Hick}]{kojima_etal:98a}
Kojima, M., Tokumaru, M., Watanabe, H., {et~al.} 1998, \jgr, 103, 1981

\bibitem[{{Kumar} {et~al.}(2018){Kumar}, {Zirnstein}, \&
  {Spitkovsky}}]{kumar_etal:18a}
{Kumar}, R., {Zirnstein}, E.~J., \& {Spitkovsky}, A. 2018, \apj, 860, 156,
  \dodoi{10.3847/1538-4357/aabf96}

\bibitem[{{Lallement} {et~al.}(2010){Lallement}, {Qu{\'e}merais}, {Lamy},
  {Bertaux}, {Ferron}, \& {Schmidt}}]{lallement_etal:10b}
{Lallement}, R., {Qu{\'e}merais}, E., {Lamy}, P., {et~al.} 2010, in
  Astronomical Society of the Pacific Conference Series, Vol. 428, SOHO-23:
  Understanding a Peculiar Solar Minimum, ed. {S.~R.~Cranmer, J.~T.~Hoeksema,
  \& J.~L.~Kohl}, 253--258.
\newblock \doarXiv{1003.4243}

\bibitem[{{Le Chat} {et~al.}(2012){Le Chat}, {Issautier}, \&
  {Meyer-Vernet}}]{leChat_etal:12a}
{Le Chat}, G., {Issautier}, K., \& {Meyer-Vernet}, N. 2012, \solphys, 279, 197,
  \dodoi{10.1007/s11207-012-9967-y}

\bibitem[{Machol {et~al.}(2019)Machol, Snow, Woodraska, Woods, Viereck, \&
  Coddington}]{machol_etal:19a}
Machol, J.~L., Snow, M., Woodraska, D., {et~al.} 2019, Earth and Space Science,
  6, 2263, \dodoi{10.1029/2019EA000648}

\bibitem[{{Manoharan}(1993)}]{manoharan:93b}
{Manoharan}, P.~K. 1993, \solphys, 148, 153, \dodoi{10.1007/BF00675541}

\bibitem[{{McComas} {et~al.}(2013){McComas}, Angold, Elliott, Livadiotis,
  Schwadron, Skoug, \& Smith}]{mccomas_etal:13b}
{McComas}, D.~J., Angold, N., Elliott, H.~A., {et~al.} 2013, \apj, 779, 2,
  \dodoi{10.1088/0004-637X/779/1/2}

\bibitem[{{McComas} {et~al.}(2008){McComas}, Ebert, Elliot, Goldstein, Gosling,
  Schwadron, \& Skoug}]{mccomas_etal:08a}
{McComas}, D.~J., Ebert, R.~W., Elliot, H.~A., {et~al.} 2008, \grl, 35, L18103,
  \dodoi{10.1029/2008GL034896}

\bibitem[{{McComas} {et~al.}(2002){McComas}, {Elliott}, {Gosling},
  {Reisenfeld}, {Skoug}, {Goldstein}, {Neugebauer}, \&
  {Balogh}}]{mccomas_etal:02b}
{McComas}, D.~J., {Elliott}, H.~A., {Gosling}, J.~T., {et~al.} 2002, \grl, 29,
  1290, \dodoi{10.1029/2001GL014164}

\bibitem[{{McComas} {et~al.}(1998){McComas}, {Bame}, {Barraclough}, {Feldman},
  {Funsten}, {Gosling}, {Riley}, {Skoug}, {Balogh}, {Forsyth}, {Goldstein}, \&
  {Neugebauer}}]{mccomas_etal:98b}
{McComas}, D.~J., {Bame}, S.~J., {Barraclough}, B.~L., {et~al.} 1998, \grl, 25,
  1, \dodoi{10.1029/97GL03444}

\bibitem[{{McComas} {et~al.}(2000){McComas}, Barraclough, Funsten, Gosling,
  Santiago-Munoz, Goldstein, Neugebauer, Riley, \& Balogh}]{mccomas_etal:00b}
{McComas}, D.~J., Barraclough, B.~L., Funsten, H.~O., {et~al.} 2000, \jgr, 105,
  10419, \dodoi{10.1029/1999JA000383}

\bibitem[{{McComas} {et~al.}(2009){McComas}, {Allegrini}, {Bochsler},
  {Bzowski}, {Christian}, {Crew}, {DeMajistre}, {Fahr}, {Fichtner}, {Frisch},
  {Funsten}, {Fuselier}, {Gloeckler}, {Gruntman}, {Heerikhuisen}, {Izmodenov},
  {Janzen}, {Knappenberger}, {Krimigis}, {Kucharek}, {Lee}, {Livadiotis},
  {Livi}, {MacDowall}, {Mitchell}, {M{\"o}bius}, {Moore}, {Pogorelov},
  {Reisenfeld}, {Roelof}, {Saul}, {Schwadron}, {Valek}, {Vanderspek}, {Wurz},
  \& {Zank}}]{mccomas_etal:09c}
{McComas}, D.~J., {Allegrini}, F., {Bochsler}, P., {et~al.} 2009, Science, 326,
  959, \dodoi{10.1126/science.1180906}

\bibitem[{{McComas} {et~al.}(2018){McComas}, {Dayeh}, {Funsten},
  {Heerikhuisen}, {Janzen}, {Reisenfeld}, {Schwadron}, {Szalay}, \&
  {Zirnstein}}]{mccomas_etal:18a}
{McComas}, D.~J., {Dayeh}, M.~A., {Funsten}, H.~O., {et~al.} 2018, \apjl, 856,
  L10, \dodoi{10.3847/2041-8213/aab611}

\bibitem[{{Mostafavi} {et~al.}(2019){Mostafavi}, {Zank}, {Zirnstein}, \&
  {McComas}}]{mostafavi_etal:19a}
{Mostafavi}, P., {Zank}, G.~P., {Zirnstein}, E.~J., \& {McComas}, D.~J. 2019,
  \apjl, 878, L24, \dodoi{10.3847/2041-8213/ab25f4}

\bibitem[{{Mousavi} {et~al.}(2022){Mousavi}, {Liu}, \&
  {Sadeghzadeh}}]{mousavi_etal:22a}
{Mousavi}, A., {Liu}, K., \& {Sadeghzadeh}, S. 2022, \mnras, 510, 1031,
  \dodoi{10.1093/mnras/stab3443}

\bibitem[{Porowski {et~al.}(2022)Porowski, Bzowski, \&
  Tokumaru}]{porowski_etal:22a}
Porowski, C., Bzowski, M., \& Tokumaru, M. 2022, \apjs, 259, 2,
  \dodoi{10.3847/1538-4365/ac35d7}

\bibitem[{{Prabhakaran Nayar}(2006)}]{prabhakaran:06a}
{Prabhakaran Nayar}, S.~R. 2006, in Proceedings of the ILWS Workshop, ed.
  N.~{Gopalswamy} \& A.~{Bhattacharyya}, 170

\bibitem[{Rahmanifard {et~al.}(2019)Rahmanifard, M{\"o}bius, Schwadron, Galli,
  Richards, Kucharek, Sok{\'{o}}{\l}, Heirtzler, Lee, Bzowski,
  Kowalska-Leszczynska, Kubiak, Wurz, Fuselier, \&
  McComas}]{rahmanifard_etal:19a}
Rahmanifard, F., M{\"o}bius, E., Schwadron, N.~A., {et~al.} 2019, \apj, 887,
  217, \dodoi{10.3847/1538-4357/ab58ce}

\bibitem[{Reisenfeld {et~al.}(2019)Reisenfeld, Bzowski, Funsten, Janzen, Karna,
  Kubiak, McComas, Schwadron, \& Sok{\'o}{\l}}]{reisenfeld_etal:19a}
Reisenfeld, D.~B., Bzowski, M., Funsten, H.~O., {et~al.} 2019, \apj, 879, 1,
  \dodoi{10.3847/1538-4357/ab22c0}

\bibitem[{Ruci{\'n}ski \& Bzowski(1995)}]{rucinski_bzowski:95b}
Ruci{\'n}ski, D., \& Bzowski, M. 1995, \aap, 296, 248

\bibitem[{{Schwadron} \& {McComas}(2010)}]{schwadron_mccomas:10a}
{Schwadron}, N.~A., \& {McComas}, D.~J. 2010, \apjl, 712, L157,
  \dodoi{10.1088/2041-8205/712/2/L157}

\bibitem[{{Shimojo} {et~al.}(2017){Shimojo}, {Iwai}, {Asai}, {Nozawa},
  {Minamidani}, \& {Saito}}]{shimojo_etal:17a}
{Shimojo}, M., {Iwai}, K., {Asai}, A., {et~al.} 2017, \apj, 848, 62,
  \dodoi{10.3847/1538-4357/aa8c75}

\bibitem[{{Shlens}(2014)}]{shlens:14a}
{Shlens}, J. 2014, arXiv e-prints, arXiv:1404.1100.
\newblock \doarXiv{1404.1100}

\bibitem[{{Snow} {et~al.}(2014){Snow}, {Weber}, {Machol}, {Viereck}, \&
  {Richard}}]{snow_etal:14a}
{Snow}, M., {Weber}, M., {Machol}, J., {Viereck}, R., \& {Richard}, E. 2014,
  Journal of Space Weather and Space Climate, 4, A04,
  \dodoi{10.1051/swsc/2014001}

\bibitem[{{Sok{\'o}{\l}} {et~al.}(2019{\natexlab{a}}){Sok{\'o}{\l}}, {Bzowski},
  \& {Tokumaru}}]{sokol_etal:19a}
{Sok{\'o}{\l}}, J.~M., {Bzowski}, M., \& {Tokumaru}, M. 2019{\natexlab{a}},
  \apj, 872, 57, \dodoi{10.3847/1538-4357/aaf737}

\bibitem[{{Sok\'{o}{\l}} {et~al.}(2013){Sok\'{o}{\l}}, {Bzowski}, {Tokumaru},
  {Fujiki}, \& {McComas}}]{sokol_etal:13a}
{Sok\'{o}{\l}}, J.~M., {Bzowski}, M., {Tokumaru}, M., {Fujiki}, K., \&
  {McComas}, D.~J. 2013, \solphys, 285, 167, \dodoi{10.1007/s11207-012-9993-9}

\bibitem[{{Sok{\'o}{\l}} {et~al.}(2021){Sok{\'o}{\l}}, {Dayeh}, {Fuselier},
  {Nicolaou}, {McComas}, \& {Zirnstein}}]{sokol_etal:21a}
{Sok{\'o}{\l}}, J.~M., {Dayeh}, M.~A., {Fuselier}, S.~A., {et~al.} 2021, \apj,
  922, 250, \dodoi{10.3847/1538-4357/ac21cd}

\bibitem[{{Sok{\'o}{\l}} {et~al.}(2019{\natexlab{b}}){Sok{\'o}{\l}}, {Kubiak},
  \& {Bzowski}}]{sokol_etal:19b}
{Sok{\'o}{\l}}, J.~M., {Kubiak}, M.~A., \& {Bzowski}, M. 2019{\natexlab{b}},
  \apj, 879, 24, \dodoi{10.3847/1538-4357/ab21c4}

\bibitem[{{Sok{\'o}{\l}} {et~al.}(2020){Sok{\'o}{\l}}, {McComas}, {Bzowski}, \&
  {Tokumaru}}]{sokol_etal:20a}
{Sok{\'o}{\l}}, J.~M., {McComas}, D.~J., {Bzowski}, M., \& {Tokumaru}, M. 2020,
  \apj, 897, 179, \dodoi{10.3847/1538-4357/ab99a4}

\bibitem[{Tanaka \& Kakinuma(1957)}]{tanaka_kakinuma:57a}
Tanaka, H., \& Kakinuma, T. 1957, Proceedings of the Research Institute of
  Atmospherics, 4, 60

\bibitem[{{Tapping}(2013)}]{tapping:13a}
{Tapping}, K.~F. 2013, Space Weather, 11, 1, \dodoi{10.1002/swe.20064}

\bibitem[{Tokumaru {et~al.}(2021)Tokumaru, Fujiki, Kojima, \&
  Iwai}]{tokumaru_etal:21a}
Tokumaru, M., Fujiki, K., Kojima, M., \& Iwai, K. 2021, \apj, 922, 73,
  \dodoi{10.3847/1538-4357/ac1862}

\bibitem[{{Tokumaru} {et~al.}(2010){Tokumaru}, {Kojima}, \&
  {Fujiki}}]{tokumaru_etal:10a}
{Tokumaru}, M., {Kojima}, M., \& {Fujiki}, K. 2010, \jgr, 115, A04102,
  \dodoi{10.1029/2009JA014628}

\bibitem[{{Tokumaru} {et~al.}(2012){Tokumaru}, {Kojima}, \&
  {Fujiki}}]{tokumaru_etal:12b}
---. 2012, \jgr, 117, 6108, \dodoi{10.1029/2011JA017379}

\bibitem[{{Veronig} {et~al.}(2021){Veronig}, {Jain}, {Podladchikova}, {Poetzi},
  \& {Clette}}]{veronig_etal:21b}
{Veronig}, A.~M., {Jain}, S., {Podladchikova}, T., {Poetzi}, W., \& {Clette},
  F. 2021, VizieR Online Data Catalog, J/A+A/652/A56

\bibitem[{Viereck {et~al.}(2004)Viereck, Floyd, Crane, Woods, Knapp, Rottman,
  Weber, Puga, \& {DeLand}}]{viereck_etal:04a}
Viereck, R., Floyd, L.~E., Crane, P.~C., {et~al.} 2004, Space Weather, 2,
  S10005, \dodoi{10.1029/1004SW000084}

\bibitem[{{Wenzel} {et~al.}(1989){Wenzel}, {Marsden}, {Page}, \&
  {Smith}}]{wenzel_etal:89a}
{Wenzel}, K.-P., {Marsden}, R.~G., {Page}, D.~E., \& {Smith}, E.~J. 1989, \asr,
  9, 25, \dodoi{10.1016/0273-1177(89)90089-6}

\bibitem[{{Yang} {et~al.}(2015){Yang}, {Liu}, {Richardson}, {Lu}, {Huang}, \&
  {Wang}}]{yang_etal:15a}
{Yang}, Z., {Liu}, Y.~D., {Richardson}, J.~D., {et~al.} 2015, \apj, 809, 28,
  \dodoi{10.1088/0004-637X/809/1/28}

\bibitem[{{Zank} {et~al.}(1996){Zank}, {Pauls}, {Cairns}, \&
  {Webb}}]{zank_etal:96b}
{Zank}, G.~P., {Pauls}, H.~L., {Cairns}, I.~H., \& {Webb}, G.~M. 1996, \jgr,
  101, 457, \dodoi{10.1029/95JA02860}

\bibitem[{{Zirnstein} {et~al.}(2016){Zirnstein}, {Funsten}, {Heerikhuisen}, \&
  {McComas}}]{zirnstein_etal:15c}
{Zirnstein}, E.~J., {Funsten}, H.~O., {Heerikhuisen}, J., \& {McComas}, D.~J.
  2016, \aap, 586, A31, \dodoi{10.1051/0004-636/201527437}

\bibitem[{{Zirnstein} {et~al.}(2015){Zirnstein}, {Heerikhuisen}, \&
  {McComas}}]{zirnstein_etal:15a}
{Zirnstein}, E.~J., {Heerikhuisen}, J., \& {McComas}, D.~J. 2015, \apjl, 804,
  L22, \dodoi{10.1088/2041-8205/804/1/L22}

\bibitem[{{Zirnstein} {et~al.}(2021){Zirnstein}, {Kumar}, {Bandyopadhyay},
  {Dayeh}, {Heerikhuisen}, \& {McComas}}]{zirnstein_etal:21c}
{Zirnstein}, E.~J., {Kumar}, R., {Bandyopadhyay}, R., {et~al.} 2021, \apjl,
  916, L21, \dodoi{10.3847/2041-8213/ac12cc}

\end{thebibliography}
\bibliographystyle{aasjournal}

\end{document}